\documentclass[journal]{IEEEtran}

\usepackage[T1]{fontenc}
\usepackage{graphicx}
\usepackage{cite}
\usepackage{caption}
\usepackage{subcaption}

\usepackage{overpic}
\usepackage{amssymb}
\usepackage{amsmath}
\usepackage{amsthm}
\usepackage{array}
\usepackage{color}
\usepackage{url}

\IEEEoverridecommandlockouts

\theoremstyle{plain}

\newtheorem{corollary}{Corollary}
\newtheorem{remark}{Remark}
\newtheorem{proposition}{Proposition}

\newcommand{\mathacr}[1]{\mathsf{#1}}

\newcommand{\vect}[1]{\mathbf{#1}}

\newcommand{\condSum}[3]{\overset{#3}{\underset{\underset{#2}{#1}}{\sum}}}

\def\diag{\mathrm{diag}}

\def\tr{\mathrm{tr}}

\def\Htran{\mbox{\tiny $\mathrm{H}$}}
\def\Ttran{\mbox{\tiny $\mathrm{T}$}}
\def\CN{\mathcal{N}_{\mathbb{C}}} %Complex Gaussian
\begin{document}

\title{\huge{Making Cell-Free Massive MIMO Competitive With \\ MMSE Processing and Centralized Implementation}}

\author{
\IEEEauthorblockN{Emil Bj{\"o}rnson, \emph{Senior Member, IEEE}, Luca Sanguinetti, \emph{Senior Member, IEEE}
\thanks{
\newline\indent A conference version of this paper was presented at IEEE SPAWC 2019 \cite{Bjornson2019b}.
\newline\indent E. Bj{\"o}rnson was supported by ELLIIT and the Wallenberg AI, Autonomous Systems and Software Program (WASP). L. Sanguinetti was supported by the University of Pisa under the PRA 2018-2019 Research Project CONCEPT.
\newline \indent E.~Bj\"ornson is with the Department of Electrical Engineering (ISY), Link\"{o}ping University, 58183 Link\"{o}ping, Sweden (emil.bjornson@liu.se). L.~Sanguinetti is with the Dipartimento di Ingegneria dell'Informazione, University of Pisa, 56122 Pisa, Italy (luca.sanguinetti@unipi.it).}
}}

\maketitle

\begin{abstract} Cell-free Massive MIMO is considered as a promising technology for satisfying the increasing number of users and high rate expectations in beyond-5G networks. The key idea is to let many distributed access points (APs) communicate with all users in the network, possibly by using joint coherent signal processing. The aim of this paper is to provide the first comprehensive analysis of this technology under different degrees of cooperation among the APs. Particularly, the uplink spectral efficiencies of four different cell-free implementations are analyzed, with spatially correlated fading and arbitrary linear processing. It turns out that it is possible to outperform conventional Cellular Massive MIMO and small cell networks by a wide margin, but only using global or local minimum mean-square error (MMSE) combining. This is in sharp contrast to the existing literature, which advocates for maximum-ratio combining. Also, we show that a centralized implementation with optimal MMSE processing not only maximizes the SE but largely reduces the fronthaul signaling compared to the standard distributed approach. This makes it the preferred way to operate Cell-free Massive MIMO networks. Non-linear decoding is also investigated and shown to bring negligible improvements.~
\end{abstract}

\begin{IEEEkeywords}
Beyond 5G MIMO, Cell-free Massive MIMO, Cellular Massive MIMO, uplink, AP cooperation, MMSE processing, fronthaul signaling, non-linear decoding, small-cell networks.
\end{IEEEkeywords}

\IEEEpeerreviewmaketitle

\section{Introduction}

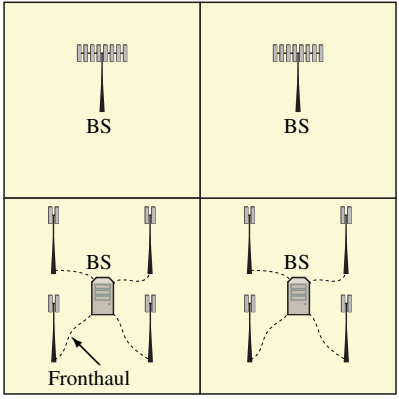
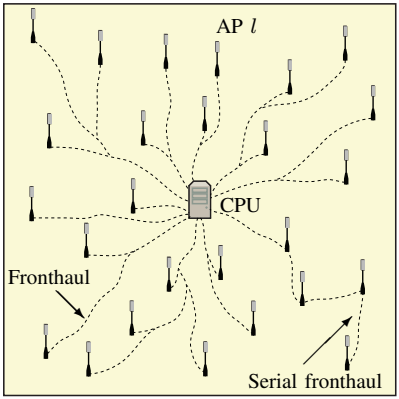
\begin{figure} 
        \centering
        \begin{subfigure}[t]{\columnwidth} \centering 
	\begin{overpic}[width=0.6\columnwidth,tics=10]{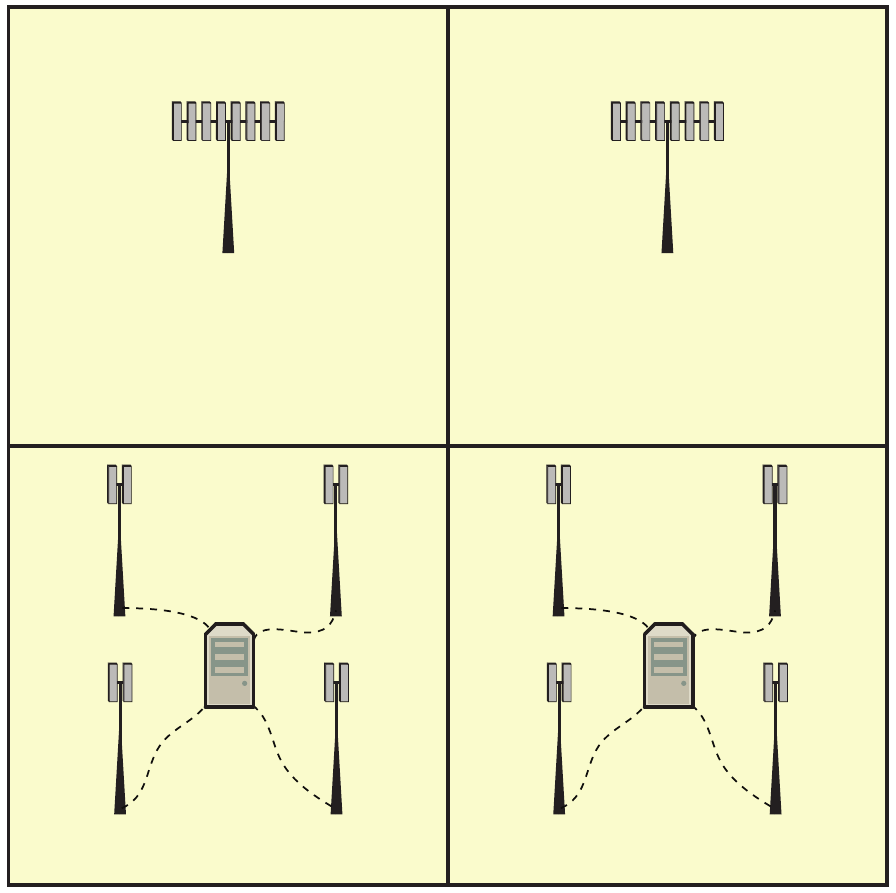}
		\put(21.5,32){\footnotesize{BS}}
		\put(71,32){\footnotesize{BS}}
		\put(21.5,66){\footnotesize{BS}}
		\put(71,66){\footnotesize{BS}}
		\put(25,8){\vector(-1,1){7}}
		\put(12,3){\footnotesize{Fronthaul}}
\end{overpic} 
                \caption{Cellular network with mMIMO BSs having either co-located arrays (top) or distributed arrays (bottom).} 
                \label{figure_SE_pathloss_iid}
        \end{subfigure} 
        \begin{subfigure}[t]{\columnwidth} \centering  \vspace{+2mm}
	\begin{overpic}[width=0.6\columnwidth,tics=10]{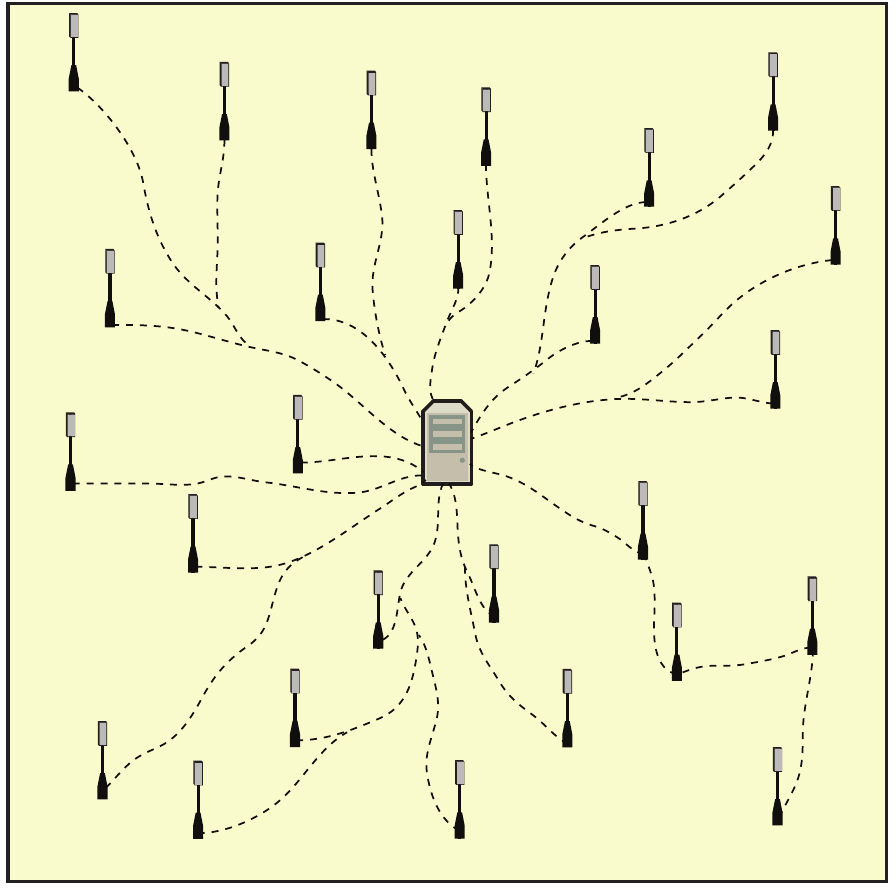}
		\put(54,92){\footnotesize{AP $l$}}
		\put(2,29){\footnotesize{Fronthaul}}
		\put(14,27.5){\vector(1,-1){7}}
		\put(62,3){\footnotesize{Serial fronthaul}}
		\put(76,9){\vector(1,1){12}}
		\put(55,47){\footnotesize{CPU}}
\end{overpic} 
                \caption{Cell-free mMIMO network.} 
                \label{figure_SE_pathloss_corr}
        \end{subfigure} 
        \caption{Comparison of different cellular and cell-free network topologies.}
        \label{fig:illustration_setups}  
\end{figure}

The traditional way to cover a large geographical area with wireless communication services uses the cellular network topology in Fig.~\ref{fig:illustration_setups}(a), where each base station (BS) serves an exclusive set of user equipments (UEs) \cite{Macdonald1979a}. This network topology has been utilized for many decades and the spectral efficiency (SE) has been gradually improved by reducing the cell sizes and applying more advanced signal processing schemes for interference mitigation \cite{massivemimobook}.

Recently, massive multiple-input multiple-output (mMIMO) has become the key 5G physical-layer technology \cite{Marzetta2010a,Larsson2014a,Andrews2014a,Parkvall2017a}. It can improve the SE by at least $10\times$ over legacy cellular networks \cite{massivemimobook}, by upgrading the BS hardware instead of deploying new BS sites. The SE gain comes from that each BS has a compact array with a hundred or more antennas, which are used for digital beamforming and, particularly, to spatially multiplex many user equipments (UEs) on the same time-frequency resource \cite{Bjornson2016a}. The characteristic feature of mMIMO, compared to traditional multi-user MIMO, is that each BS has many more antennas than UEs in the cell.
Signal processing methods, such as minimum mean-squared error (MMSE) combining in the uplink, can be used individually at each BS to suppress interference from both the same and other cells \cite{massivemimobook,BHS18A,Sanguinetti2019a}, without the need for any BS cooperation. The mMIMO theory also supports deployments with spatially distributed arrays in each cell \cite{Truong2013a,Bjornson2015b}, as also illustrated in Fig.~\ref{fig:illustration_setups}(a). This setup is essentially the same as the Distributed Antenna System (DAS) setup in \cite{Choi2007a} and Coordinated Multi-Point (CoMP) with static, disjoint cooperation clusters \cite{irmer2011coordinated,Bjornson2013d}. These are all different embodiments of cellular networks.

An alternative network infrastructure was considered in \cite{Ngo2017b,Nayebi2017a} 
under the name of \emph{Cell-free mMIMO}. The idea is to deploy a large number of distributed single-antenna access points (APs), which are connected to a central processing unit (CPU), also known as an edge-cloud processor \cite{Burr2018a} or C-RAN (cloud radio access network) data center \cite{Perlman2015a}. The CPU operates the system in a Network MIMO fashion, with no cell boundaries, to jointly serve the UEs by coherent joint transmission and reception \cite{Shamai2001a,Zhou2003a,Venkatesan2007a,Bjornson2010c,Bjornson2013d}. 
Compared to traditional Network MIMO, the outstanding aspect of Cell-free mMIMO is the operating regime with many more APs than UEs \cite{Ngo2017b}. From an analytical perspective, an important novelty was that imperfect channel state information (CSI) was considered in the performance analysis, while perfect CSI was often assumed in the past \cite{Bjornson2013d}. The paper \cite{Ngo2017b} advocated the use of maximum ratio (MR) processing (a.k.a.~matched filtering or conjugate beamforming) locally at each AP, while \cite{Nayebi2016a,Nayebi2017a} showed that partially or fully centralized processing at the CPU can achieve higher SE.

\subsection{Motivation}

The focus in the early papers \cite{Ngo2017b,Nayebi2017a} was on comparing Cell-free mMIMO with a small-cell network; that is, the APs are deployed at the same places, but each AP serves its own exclusive set of UEs. Since small cells are a special case of Cell-free mMIMO, they obviously provide lower performance. Particularly, \cite{Ngo2017b,Nayebi2017a} demonstrated large improvements in median and 95\%-likely SE. In Section~\ref{sec:numerical}, we will show that this is partially due to the fact that a poor implementation of the small-cell network was considered in \cite{Ngo2017b,Nayebi2017a}. In fact, we will show that more sophisticated processing than MR is needed in Cell-free mMIMO to always outperform small cells.

Unlike \cite{Ngo2017b,Nayebi2017a}, this paper aims at comparing Cell-free mMIMO with conventional Cellular mMIMO and its primary goal is to find the most competitive cell-free implementation.\footnote{Previous comparisons are found in \cite{Ngo2018a,Yang2018a} but only for a single cell, so it is not cellular, and only MR is used, which is known to perform badly \cite{massivemimobook}.} Both network topologies are illustrated in Fig.~\ref{fig:illustration_setups}. The large differences make the comparison non-trivial and provide interesting inputs into the design of beyond-5G networks. Cellular mMIMO benefits from channel hardening and spatial interference suppression, but cell-edge UEs can have bad channel conditions. On the other hand, Cell-free mMIMO benefits from strong macro diversity but its interference suppression capability highly depends on how it is operated. The early papers  \cite{Ngo2017b,Nayebi2017a} conjectured that channel hardening also appears in Cell-free mMIMO, but it was later shown that capacity bounds that presumes hardening can greatly underestimate the practical SE \cite{Chen2018b}. 
To achieve a reasonably fair comparison, we focus on the uplink and assume that the data transmission is preceded by a pilot-based channel estimation phase. All UEs transmit with equal powers for any of the different levels and network topologies.

\subsection{Contributions} 
The major contributions of this paper are two-fold. Firstly, we introduce a taxonomy with four different implementations of Cell-free mMIMO, which are characterized by different degrees of cooperation among the APs. Secondly, we provide new achievable SE expressions, which are valid for spatially correlated fading channels, imperfect CSI, APs with an arbitrary number $N$ of antennas, and heuristic or optimized receive combining schemes. All this provides a common analytical framework to numerically evaluate the benefits and costs (in terms of fronthaul signaling) of the different implementations and to understand how Cell-free mMIMO should be operated and designed in order to get much higher performance than conventional Cellular mMIMO and small cells.

The four different levels of cooperation that we consider in this paper are as follows. The so-called \emph{Level 4} is a form of Network MIMO and stands for a fully centralized network in which the pilot and data signals received at all APs are gathered (through the fronthaul links) at the CPU, which performs channel estimation and data detection. \emph{Level 3} relies on the large-scale fading decoding (LSFD) strategy, which was originally proposed for Cellular mMIMO in \cite{ashikhmin2012pilot,Adhikary2017a}. Particularly, it operates in two stages. In the first stage, each AP locally estimates the channels and applies an arbitrary receive combiner to obtain local estimates of the UE data. These are then gathered at the CPU where they are linearly processed to perform joint detection. Only channel statistics can be utilized in the second stage at the CPU since the pilot signals are not shared over the fronthaul links. \emph{Level 2} is a direct simplification of Level 3 in the sense that the CPU performs detection in the second stage by simply taking the average of the local estimates. This dispenses the CPU from knowledge of the channel statistics and thus reduces the amount of information to be exchanged. Finally, \emph{Level 1} stands for a fully distributed network in which the detection is performed locally at the APs by using only local channel estimates and one AP serves each UE. This is a small-cell network where nothing is exchanged with the CPU.

The above levels have been partially analyzed before in the literature, but not under the general and practical conditions considered in this paper, which allow us to draw conclusions that differ in several important ways---in particular, we show that MR combining performs terribly bad in Cell-free mMIMO.
Level 4 was considered in \cite{Nayebi2016a,Riera2018a,Yang2019a} for $N=1$ and in \cite{Chen2018b,Bashar2018a} with $N \geq 1$ but with spatially uncorrelated channels.
Level 3 was investigated in \cite{Nayebi2016a,Ngo2018b} for $N=1$ and MR combining.
Level 2 was considered in \cite{Ngo2017b,Buzzi2017a,Zhang2018a,Ozdogan2018a,Fan2019a,Bashar2019a,Yang2019a} (among many others) but only with MR combining. A suboptimal implementation of Level 1 with $N=1$ was considered in \cite{Ngo2017b} (the suboptimality is explained in detail in Section~\ref{subsec:level1}). There are also previous papers that consider various forms of Levels 1--4 under perfect CSI; see the reference list of \cite{Ngo2017b} for a good selection of such papers. In addition, there is previous research on BS cooperation in cellular networks, where the received signals and CSI are shared between BSs to cancel inter-cell interference; see \cite{Sanderovich2007a,Marsch2011c,Simeone2012a} and reference therein. These papers also consider different levels of cooperation, but these are heavily influenced by the cellular topology (e.g., BSs send signals to each other, BSs are surrounded by  UEs, and there exist cell edges) and can thus not be applied to Cell-free mMIMO.

\subsection{Paper Outline}
The rest of this paper is organized as follows. Section~\ref{sec:network-model} defines the system model for uplink Cell-free mMIMO for both data transmission and channel estimation. Next, Section~\ref{sec:four-level-receiver-cooperation} presents the four levels of receiver cooperation, including achievable SE expressions for spatially correlated fading, multi-antenna APs, and optimized receive combining. The four levels are numerically compared with Cellular mMIMO in Section~\ref{sec:numerical}. This section also discusses the differences and similarities with the previous results in \cite{Ngo2017b}. Section~\ref{sec:MMSE-SIC} evaluates the potential benefit of using non-linear decoding at the CPU, whereas the fronthaul signaling required with the different implementations is quantified in Section~\ref{sec:fronthaul}. Finally, the major conclusions and implications are drawn in Section~\ref{sec:conclusion}. 

\textit{\textbf{Reproducible research:}} All the simulation results can be reproduced using the Matlab code and data files available at:\\ \url{https://github.com/emilbjornson/competitive-cell-free}

\textit{\textbf{Notation:}} 
Boldface lowercase letters, $\vect{x}$, denote column vectors and boldface uppercase letters, $\vect{X}$, denote matrices. 
The superscripts $^{\Ttran}$, $^*$ and $^{\Htran}$ denote transpose, conjugate, and conjugate transpose, respectively. 
 The $n \times n$ identity matrix is $\vect{I}_n$.
We use $\triangleq$ for definitions and $\diag(\vect{A}_1,\ldots,\vect{A}_n)$ for a block-diagonal matrix with the square matrices $\vect{A}_1,\ldots,\vect{A}_n$ on the diagonal.
The multi-variate circularly symmetric complex Gaussian distribution with correlation matrix $\vect{R}$ is denoted $\CN(\vect{0},\vect{R})$. The expected value of $\vect{x}$ is denoted as $\mathbb{E}\{ \vect{x} \}$.

\section{Cell-Free Network Model} \label{sec:network-model}

We consider a Cell-free mMIMO network consisting of $L$ geographically distributed APs, each equipped with $N$ antennas. The APs are connected via fronthaul connections to a CPU, as illustrated in Fig.~\ref{fig:illustration_setups}(b). There are $K$ single-antenna UEs in the network and the channel between AP $l$ and UE $k$ is denoted by $\vect{h}_{kl} \in \mathbb{C}^N$. We use the standard block fading model where $\vect{h}_{kl}$ is constant in time-frequency blocks of $\tau_c$ channel uses \cite{massivemimobook}. In each block, an independent realization from a correlated Rayleigh fading distribution is drawn:
\begin{equation}
\vect{h}_{kl} \sim \CN(\vect{0}, \vect{R}_{kl})
\end{equation}
where $\vect{R}_{kl} \in \mathbb{C}^{N \times N}$ is the spatial correlation matrix, which describes the spatial properties of the channel and $\beta_{kl} \triangleq \tr(\vect{R}_{kl})/N$ is the large-scale fading coefficient that describes geometric pathloss and shadowing. 

This paper considers the uplink, which consists of $\tau_p $ channel uses dedicated for pilots and $\tau_c - \tau_p$ channel uses for payload data. The two phases are described below. Notice that the results of this paper apply to both systems operating in time-division duplex (TDD) and frequency-division duplex (FDD) mode, since the uplink works the same in both cases.

\subsection{Pilot Transmission and Channel Estimation}

We assume that $\tau_p$ mutually orthogonal $\tau_p$-length pilot signals $\boldsymbol{\phi}_{1},\ldots,\boldsymbol{\phi}_{\tau_p}$ with $\| \boldsymbol{\phi}_{t} \|^2 = \tau_p$ are used for channel estimation. These pilots are assigned to the UEs in a deterministic but arbitrary way. The case of practical interest is a large network with $K > \tau_p$ so that more than one UE is assigned to each pilot. 
We denote  the index of the pilot assigned to UE $k$ as $t_k \in \{ 1, \ldots, \tau_p\}$ and call $\mathcal{P}_k \subset \{ 1, \ldots, K\}$ the subset of UEs that use the same pilot as UE $k$, including itself.

When the UEs transmit their pilots, the received signal $\vect{Z}_{l} \in \mathbb{C}^{N \times \tau_p}$ at AP $l$ is 
\begin{equation} \label{eq:received-pilot-matrix}
\vect{Z}_{l} = \sum_{i=1}^{K} \sqrt{p_i } \vect{h}_{il} \boldsymbol{\phi}_{t_i}^{\Ttran}+ \vect{N}_{l}
\end{equation}
where $p_i \geq 0$ is the transmit power of UE $i$, $\vect{N}_{l}  \in \mathbb{C}^{N \times \tau_p}$ is the receiver noise with independent $ \CN (0, \sigma^2 )$ entries, and  $\sigma^2$ is the noise power.
To estimate $\vect{h}_{kl}$, the AP first correlates the received signal with the associated normalized pilot signal $\boldsymbol{\phi}_{t_k}/ \sqrt{\tau_p}$ to obtain $\vect{z}_{t_k l} \triangleq \frac{1}{\sqrt{\tau_p}} \vect{Z}_{l}  \boldsymbol{\phi}_{t_k}^{*} \in \mathbb{C}^N$, which is given by
\begin{align} \notag
 \vect{z}_{t_k l} & = 
 \sum_{i=1}^{K}  \frac{\sqrt{p_i }}{\sqrt{\tau_p}}\vect{h}_{il} \boldsymbol{\phi}_{t_i}^{\Ttran}  \boldsymbol{\phi}_{t_k}^{*}+ \frac{1}{\sqrt{\tau_p}} \vect{N}_{l}  \boldsymbol{\phi}_{t_k}^{*}  \\
 & = \sum_{i \in \mathcal{P}_k} \sqrt{p_i \tau_p } \vect{h}_{il} + \vect{n}_{t_k l}
 \label{eq:received-pilot}
\end{align}
where $\vect{n}_{t_k l} \triangleq  \vect{N}_{l}  \boldsymbol{\phi}_{t_k}^{*}/ \sqrt{\tau_p} \sim \CN (\vect{0}, \sigma^2 \vect{I}_N)$ is the resulting noise. Using standard results from estimation theory \cite[Sec.~3]{massivemimobook},
the MMSE estimate of $\vect{h}_{kl}$ is
\begin{equation} \label{eq:estimates}
\hat{\vect{h}}_{kl} = \sqrt{p_k \tau_p} \vect{R}_{kl} \vect{\Psi}_{t_kl}^{-1} \vect{z}_{t_kl}  
\end{equation}
where
\begin{equation} \label{eq:Psitl}
\vect{\Psi}_{t_kl} = \mathbb{E} \{ \vect{z}_{t_k l} \vect{z}_{t_k l}^{\Htran} \} = \sum_{i \in \mathcal{P}_k} \tau_p p_i \vect{R}_{il} + \vect{I}_{N}
\end{equation}
is the correlation matrix of the received signal in \eqref{eq:received-pilot}.
The estimate $\hat{\vect{h}}_{kl}$ and estimation error $\tilde{\vect{h}}_{kl} = \vect{h}_{kl} - \hat{\vect{h}}_{kl}$ are independent vectors distributed as $\hat{\vect{h}}_{kl}\sim \CN \left( \vect{0}, p_k \tau_p \vect{R}_{kl} \vect{\Psi}_{t_kl}^{-1} \vect{R}_{kl} \right)$ and $\tilde{\vect{h}}_{kl}\sim \CN(\vect{0},\vect{C}_{kl})$ with
\begin{equation}
\vect{C}_{kl} = \mathbb{E} \{ \tilde{\vect{h}}_{kl} \tilde{\vect{h}}_{kl}^{\Htran} \}= \vect{R}_{kl} - p_k \tau_p \vect{R}_{kl} \vect{\Psi}_{t_kl}^{-1} \vect{R}_{kl}.
\end{equation} 
The mutual interference generated by the pilot-sharing UEs in \eqref{eq:received-pilot} causes the so-called \emph{pilot contamination} that degrades the system performance, similar to the case in Cellular mMIMO.

\begin{remark} \label{remark:channel-estimation}
The computation of $\hat{\vect{h}}_{kl}$ in \eqref{eq:estimates} requires knowledge of the correlation matrices $\{\vect{R}_{il}:i\in \mathcal{P}_k\}$, which we assume to be locally available at AP $l$; see \cite{massivemimobook} for methods to estimate them. To dispense with their full knowledge, the AP can apply alternative channel estimation schemes as in Cellular mMIMO \cite[Sec.~3.4]{massivemimobook}. One option is the so-called element-wise MMSE estimator that uses only the main diagonals of $\{\vect{R}_{il}:i\in \mathcal{P}_k\}$. Alternatively, the least-square estimator can be used, which requires no prior statistical information and computes the estimate of ${\vect{h}}_{kl}$ as $\hat{\vect{h}}_{kl}=\frac{1}{\sqrt{p_k \tau_p }}\vect{z}_{t_kl}$; see \cite{Fan2019a}.
\end{remark}

\subsection{Uplink Data Transmission}

During the uplink data transmission, the received complex baseband signal $\vect{y}_{l} \in \mathbb{C}^{N}$ at AP $l$ is given by
\begin{equation} \label{eq:received-data}
\vect{y}_{l} = \sum_{i=1}^{K} \vect{h}_{il} s_i + \vect{n}_{l}
\end{equation}
where $s_i\sim\CN(0,p_i)$ is the information-bearing signal transmitted by UE~$i$ with power $p_i$ and $\vect{n}_{l}\sim \CN (\vect{0}, \sigma^2 \vect{I}_N)$
is the independent receiver noise. 

\begin{remark}
The signal model in \eqref{eq:received-pilot-matrix} and \eqref{eq:received-data} implicitly assumes that the entire network is synchronized in time. There exist wired and over-the-air methods that can be used to synchronize the clocks at the APs \cite{Rogalin2014a,Perlman2015a,Interdonato2018}. However, the signal transmitted by a UE will never be synchronously received by all the APs due to the largely different distances between the UE and different APs.
In orthogonal frequency-division multiplexing systems, a simple way to compensate for that is to select the length of the cyclic prefix so as to accommodate both the channel delay spread and timing misalignments. This results in a quasi-synchronous system \cite{Morelli2007}. For example, in the LTE standard, the cyclic prefix is long enough to assume that a UE is quasi-synchronized to all APs within a 1\,km radius. If the extended cyclic prefix is used, the range increases up to 5\,km. Since the APs that are further away will receive negligible signal power, the model in \eqref{eq:received-pilot-matrix} and \eqref{eq:received-data} is accurate enough for the performance analysis considered in this paper.
\end{remark}

\section{Four Levels of Receiver Cooperation} \label{sec:four-level-receiver-cooperation}

All the APs are connected via fronthaul connections to a CPU that has high computational resources.\footnote{In practice, cell-free systems will have more than one CPU and only a subset of the APs will serve each UE \cite{Bjornson2013d,Perlman2015a,Interdonato2018}. The methods described in this paper applies also to that case. The only requirement is that each UE is assigned to one CPU that takes partial or full responsibility for the decoding of the UE's data and will then forward the decoded data to the core network.} 
Hence, the APs can be viewed as remote-radio heads that cooperate to support coherent communication with the UEs. 
The fronthaul can consist of a mix of wired and wireless connections, organized in a star or mesh topology \cite{Perlman2015a}; the methods developed in this paper can be applied with any fronthaul topology.
AP $l$ receives the signal $\vect{y}_{l} $ in \eqref{eq:received-data} and can use the available channel estimates $\{ \hat{\vect{h}}_{kl} : k=1,\ldots,K \}$ to detect the data signals locally, or can fully or partially delegate this task to the CPU. The benefit of using the CPU is that it can combine the inputs from all APs, but this must be balanced against the required amount of fronthaul signaling. Four levels of receiver cooperation are described below and compared with Cellular mMIMO in Section~\ref{sec:numerical} by means of numerical results.

\subsection{Level 4: Fully Centralized Processing}

The most advanced level of Cell-free mMIMO operation is when the $L$ APs send their received pilot signals $\{ \vect{z}_{t l} : t=1,\ldots,\tau_p, l=1,\ldots,L \}$ and received data signals $\{ \vect{y}_{l} : l=1,\ldots, L \}$ to the CPU, which takes care of the channel estimation and data signal detection. In other words, the APs act as relays that forward all signals to the CPU \cite{Estella2019a}. In each coherence block, each AP needs to send $\tau_p N$ complex scalars for the pilot signals and $(\tau_c-\tau_p)N$ complex scalars for the received signals. This becomes $\tau_c N$ complex scalars in total, which is summarized in Table~\ref{tab:signaling}. Moreover, the spatial correlation matrices $\{  \vect{R}_{kl} : k=1,\ldots,K, l=1,\ldots,L\}$ are assumed available at the CPU at Level 4, which are described by $KLN^2$ real scalars or $KLN^2/2$ complex scalars.\footnote{It is not strictly necessary for the CPU to know the spatial correlation matrices, but it can use estimators that do not require that; see Remark~\ref{remark:channel-estimation}.}

The received signal available at the CPU is expressed as
\begin{equation} \label{eq:received-data-central}
\underbrace{ \begin{bmatrix} \vect{y}_{1} \\ \vdots \\ \vect{y}_{L} 
\end{bmatrix}}_{\triangleq \vect{y}} =   \sum_{i=1}^{K} \underbrace{\begin{bmatrix} \vect{h}_{i1} \\ \vdots \\ \vect{h}_{iL} 
\end{bmatrix}}_{\triangleq \vect{h}_i}  s_i + \underbrace{\begin{bmatrix} \vect{n}_{1} \\ \vdots \\ \vect{n}_{L} 
\end{bmatrix}}_{\triangleq \vect{n}}
\end{equation}
or, in a more compact form, as 
\begin{equation} \label{eq:received-data-central2}
\vect{y} = \sum_{i=1}^{K} \vect{h}_{i} s_i + \vect{n}.
\end{equation}
The collective channel is distributed as $\vect{h}_k \sim \CN(\vect{0}, \vect{R}_{k})$ where $\vect{R}_{k} = \diag(\vect{R}_{k1}, \ldots,  \vect{R}_{kL}) \in \mathbb{C}^{LN \times LN}$ is the block-diagonal spatial correlation matrix. Notice that \eqref{eq:received-data-central2} is mathematically equivalent to the signal model of a single-cell mMIMO system with correlated fading \cite[Sec.~2.3.1]{massivemimobook}. The only difference is how the correlation matrices are generated and how the pilots are allocated. In fact, in conventional single-cell mMIMO orthogonal pilots are assigned to UEs whereas the same pilot can be assigned to multiple UEs in the investigated cell-free network. This leads to pilot contamination between UEs served by the same AP antennas.

\begin{table}[t]  
        \caption{Number of complex scalars to send from the APs to the CPU via the fronthaul, either in each coherence block or for each realization of the user locations/statistics.}  \label{tab:signaling}
\centering
    \begin{tabular}{|c|c|c|} 
    \hline
    {} & {Each coherence block}&{Statistical parameters}   \\      \hline\hline 

   Level 4 &   $\tau_c NL $  &  $KLN^2/2$    \\ \hline    
   Level 3 &   $(\tau_c-\tau_p) KL $ & $KL+(L^2 K^2+KL)/2$  \\   \hline    
   Level 2 &   $(\tau_c-\tau_p) KL $ & $-$   \\   \hline    
   Level 1 &   $-$ & $-$ 
   \\ \hline
    \end{tabular}
    \vspace{-0.2cm}
\end{table}

The CPU can compute all the MMSE channel estimates $\{\hat{\vect{h}}_{il} : k=1,\ldots,K, l=1,\ldots,L\}$ using the received pilot signals and channel statistics obtained from the APs. The estimates can be computed separately without loss of optimality. For UE $k$, the CPU can then form the collective channel estimate
\begin{equation} \label{eq:estimates-central}
\hat{\vect{h}}_{k} \triangleq \begin{bmatrix} \hat{\vect{h}}_{k1} \\ \vdots \\ \hat{\vect{h}}_{kL} 
\end{bmatrix} \sim \CN \left( \vect{0}, p_k \tau_p \vect{R}_{k} \vect{\Psi}_{t_k}^{-1} \vect{R}_{k} \right)
\end{equation}
where $\vect{\Psi}_{t_k}^{-1} = \diag( \vect{\Psi}_{t_k1}^{-1}, \ldots, \vect{\Psi}_{t_kL}^{-1})$. The estimation error is $\tilde{\vect{h}}_k = \vect{h}_k - \hat{\vect{h}}_{k} \sim \CN(\vect{0},\vect{C}_k)$ with
$\vect{C}_k = \diag(\vect{C}_{k1}, \ldots, \vect{C}_{kL})$.
Next, the CPU selects an arbitrary receive combining vector $\vect{v}_k \in \mathbb{C}^{LN}$ for UE $k$ based on all the collective channel estimates $\{ \hat{\vect{h}}_{k} : k=1,\ldots,K\}$.

While the capacity of Level 4 networks with perfect CSI is known in some cases \cite{Estella2019a}, the ergodic capacity is generally unknown in the considered case with imperfect CSI. However, we can rigorously analyze the performance by using standard capacity lower bounds
\cite{Biglieri1998a,massivemimobook}, which we refer to as achievable SEs.

\begin{proposition} \label{theorem:uplink-capacity-general}
At Level 4, if the MMSE estimator is used to compute channel estimates for all UEs, an achievable SE of UE $k$ is
\begin{equation} \label{eq:uplink-rate-expression-general}
\mathacr{SE}_{k}^{(4)} = \left( 1 - \frac{\tau_p}{\tau_c} \right) \mathbb{E} \left\{ \log_2  \left( 1 + \mathacr{SINR}_{k}^{(4)}  \right) \right\}
\end{equation}
where the instantaneous effective signal-to-interference-and-noise ratio (SINR) is
\begin{equation} \label{eq:uplink-instant-SINR}
\!\mathacr{SINR}_{k}^{(4)} \!=  \!\frac{ p_{k} |  \vect{v}_{k}^{\Htran} \hat{\vect{h}}_{k} |^2  }{ 
\sum\limits_{i=1,i\ne k}^K p_{i} | \vect{v}_{k}^{\Htran} \hat{\vect{h}}_{i} |^2
+ \vect{v}_{k}^{\Htran}  \left( \sum\limits_{i=1}^{K} p_{i} \vect{C}_{i} + \sigma^2  \vect{I}_{LN} \right)  \vect{v}_{k}  
}\!\!\!
\end{equation}
and the expectation is with respect to the channel estimates.
\end{proposition}
\begin{IEEEproof}
The proof follows the same steps as the proof of \cite[Th.~4.1]{massivemimobook} for Cellular mMIMO and is therefore omitted.
\end{IEEEproof}

The pre-log factor $1 - {\tau_p}/{\tau_c}$ in \eqref{eq:uplink-rate-expression-general} is the fraction of channel uses that are used for uplink data transmission. The term $\mathacr{SINR}_{k}^{(4)}$ takes the form of an ``effective instantaneous SINR'' \cite{massivemimobook}, with the desired signal power received over the estimated channel in the numerator and the interference plus noise in the denominator.\footnote{The word ``effective'' refers to the fact that $\mathacr{SINR}_{k}^{(4)}$ cannot be measured in the system at any particular point in time, but the SE is the same as that of a fading single-antenna point-to-point channel where $\mathacr{SINR}_{k}^{(4)}$ is the instantaneously measurable SINR and the receiver has perfect CSI.}

We notice that the SE expression in \eqref{eq:uplink-rate-expression-general} holds for any receive combining vector $\vect{v}_{k}$ and is a multi-antenna generalization of \cite[Eq.~(1)]{Nayebi2016a} and an extension of\cite{Chen2018b,Bashar2018a} to spatially correlated channels. The expression can be easily computed for any $\vect{v}_{k}$ by using Monte Carlo methods, as done in Section~\ref{sec:numerical}. A possible choice is to use the simple MR combining with $\vect{v}_k = \hat{\vect{h}}_k$, which has low computational complexity and maximizes the power of the desired signal, but neglects the existence of interference. Other heuristic combiners such as zero-forcing (ZF) or regularized zero-forcing (RZF) can be also applied. Instead of resorting to heuristics, we notice that $\mathacr{SINR}_{k}^{(4)}$ in \eqref{eq:uplink-instant-SINR} only depends on $\vect{v}_k$ and has the form of a generalized Rayleigh quotient. Hence, the combining vector that maximizes \eqref{eq:uplink-instant-SINR} can be obtained as follows.

\begin{corollary} \label{cor:MMSE-combining}
The instantaneous SINR in \eqref{eq:uplink-instant-SINR} for UE~$k$ is maximized by the MMSE combining vector
\begin{equation} \label{eq:MMSE-combining}
\vect{v}_{k} =  p_{k}  \left( \sum\limits_{i=1}^{K} p_{i} \left( \hat{\vect{h}}_{i} \hat{\vect{h}}_{i}^{\Htran} + \vect{C}_{i} \right) + \sigma^2  \vect{I}_{LN} \right)^{-1}     \hat{\vect{h}}_{k}
\end{equation}
which leads to the maximum value
\begin{align} \label{eq:uplink-maximized-SINR}
\!\!\!\mathacr{SINR}_{k}^{(4)} \!= p_{k}  \hat{\vect{h}}_{k}^{\Htran} \!\! \left( \sum\limits_{i=1,i\ne k}^{K} p_{i}  \hat{\vect{h}}_{i} \hat{\vect{h}}_{i}^{\Htran} + \sum\limits_{i=1}^{K} p_{i} \vect{C}_{i} + \sigma^2  \vect{I}_{LN} \!\!\right)^{ \!\!-1}  \!\!\!\!\!  \hat{\vect{h}}_{k}.
\end{align}
\end{corollary}
\begin{IEEEproof}
It follows from \cite[Lemma B.10]{massivemimobook} since \eqref{eq:uplink-rate-expression-general} is a generalized Rayleigh quotient with respect to $\vect{v}_{k}$.
\end{IEEEproof}

It can be shown that the SINR-maximizing combiner in \eqref{eq:MMSE-combining} minimizes the mean-squared error ${\rm{MSE}}_k = \mathbb{E} \{ | s_{k} - \vect{v}_{k}^{\Htran} \vect{y}  |^2  \big| \{ \hat{\vect{h}}_{i} \}  \}$, which represents the conditional MSE between the data signal $s_k$ and the received signal $\vect{v}_{k}^{\Htran} \vect{y}$ after
receive combining; see \cite[Sec.~4.1]{massivemimobook} for details. This is why it is called \emph{MMSE combining}. This type of receive combining normally maximizes the mutual information of channels with multiple receive antennas \cite{Park2013a}, but the particular expression in \eqref{eq:MMSE-combining} is unique for Cell-free mMIMO.

Compared to many heuristic solutions, MMSE combining has higher computational complexity since it requires first the computation of the $LN\times LN$ matrix inverse in \eqref{eq:MMSE-combining} and then a matrix-vector multiplication. However, this is not a major issue since it has to be implemented at the CPU, which is assumed to have high computational capability.
If the complexity is a concern, then ZF and RZF can be used instead since only $K \times K$ matrices need to be inverted. The price to pay is that the UEs with low SNRs get an SE reduction, which may be very large.

\subsection{Level 3: Local Processing \&  Large-Scale Fading Decoding}

Instead of sending the $N$-dimensional vectors $\{ \vect{y}_{l} : l=1,\ldots, L \}$ and the channel estimates to the CPU, each AP can preprocess its  signal by computing local estimates of the data that are then passed to the CPU for final decoding. Let $\vect{v}_{kl} \in \mathbb{C}^{N}$ be the local combining vector that AP~$l$ selects for UE~$k$. Then, its local estimate of $s_k$ is
\begin{equation} \label{eq:local-data-estimate}
\check{s}_{kl} \triangleq \vect{v}_{kl}^{\Htran} \vect{y}_{l} = \vect{v}_{kl}^{\Htran}\vect{h}_{kl} s_k +  \sum_{i=1,i\ne k}^{K} \vect{v}_{kl}^{\Htran} \vect{h}_{il} s_i + \vect{v}_{kl}^{\Htran}\vect{n}_{l}.
\end{equation}
Any combining vector can be adopted in the above expression. Unlike at Level 4, however, AP $l$ can only use its own local channel estimates $\{\hat{\bf h}_{il}: i=1,\ldots,K\}$ for the design of $\vect{v}_{kl}$. The simplest solution is MR combining with $\vect{v}_{kl} = \hat{\vect{h}}_{kl}$ as in \cite{Ngo2017b,Nayebi2016a} but preferably the AP should use its local CSI to make $\check{s}_{kl}$ as close to $s_k$ as possible.
The combining vector that minimizes the MSE, ${\rm{MSE}}_{kl} =\mathbb{E} \{ | s_{k} - \vect{v}_{kl}^{\Htran} \vect{y}_{l} |^2  \big| \{ \hat{\vect{h}}_{il} \}  \}$, is
\begin{equation} \label{eq:MMSE-combining-single-AP}
\vect{v}_{kl} =  p_{k}  \left( \sum\limits_{i=1}^{K} p_{i} \left( \hat{\vect{h}}_{il} \hat{\vect{h}}_{il}^{\Htran} + \vect{C}_{il} \right) + \sigma^2  \vect{I}_{N} \right)^{-1}     \hat{\vect{h}}_{kl}
\end{equation}
which can be proved by computing the conditional expectation and equating the first derivative with respect to $\vect{v}_{kl}$ to zero. Notice that \eqref{eq:MMSE-combining-single-AP} is the combining vector that would maximize the SE if AP $l$ decoded the data signal $s_{k}$ locally. 
We call \eqref{eq:MMSE-combining-single-AP} \emph{Local MMSE (L-MMSE) combining} to distinguish it from the MMSE combining in \eqref{eq:MMSE-combining} at Level 4, which is applied at the CPU. A main benefit over MMSE combining is that an $N \times N$ matrix is inverted in \eqref{eq:MMSE-combining-single-AP} instead of an $LN \times LN$ matrix. Importantly, even if $N=1$, \eqref{eq:MMSE-combining-single-AP} is not equal to MR but differ by a non-deterministic scaling factor.

The local estimates $\{\check{s}_{kl}: l=1,\ldots,L\}$ are then sent to the CPU where they are linearly combined using the weights $\{a_{kl}: l=1,\ldots,L\}$ to obtain $\hat{s}_k = \sum_{l=1}^{L} a_{kl}^* \check{s}_{kl}$, which is eventually used to decode $s_{k}$. From \eqref{eq:local-data-estimate}, we have that 
\begin{align} 
\hat{s}_k =\left(\sum_{l=1}^{L} a_{kl}^* \vect{v}_{kl}^{\Htran} \vect{h}_{kl}\right)s_k \!+ \!  \sum_{l=1}^{L} a_{kl}^* \!\Bigg(\sum\limits_{i=1,i\ne k}^{K}\vect{v}_{kl}^{\Htran} \vect{h}_{il} s_i\!\!\Bigg) + \vect{n}^\prime_{k} \!\!\label{eq:CPU_linearcomb}\!\!\!\!
\end{align}
with $\vect{n}^\prime_{k} = \sum_{l=1}^{L} a_{kl}^* \vect{v}_{kl}^{\Htran}\vect{n}_{l}$. Let $\vect{g}_{ki} = [ \vect{v}_{k1}^{\Htran} \vect{h}_{i1} \, \ldots \, \vect{v}_{kL}^{\Htran} \vect{h}_{iL}]^{\Ttran}$ be the $L$-dimensional vector with the receive-combined channels between UE $k$ and each of the APs. Then, \eqref{eq:CPU_linearcomb} reduces to 
\begin{align} 
\hat{s}_k =\vect{a}_{k}^{\Htran}\vect{g}_{kk}s_k + \sum\limits_{i=1,i\ne k}^{K}\vect{a}_{k}^{\Htran}\vect{g}_{ki} s_i + \vect{n}^\prime_{k} \!\!\label{eq:CPU_linearcomb_eff}
\end{align}
where $\vect{a}_{k} = [ a_{k1} \, \ldots \, a_{kL} ]^{\Ttran} \in \mathbb{C}^L$ is the weighting coefficient vector and $\{\vect{a}_{k}^{\Htran}\vect{g}_{ki}: i=1,\ldots,K\}$ represent the effective channels. Notice that $\vect{a}_{k}$ can be optimized by the CPU to maximize the SE, but only channel statistics can be utilized since the CPU does not have knowledge of the channel estimates at Level 3. This approach is known as LSFD in Cellular mMIMO \cite{ashikhmin2012pilot,Adhikary2017a}, and can be applied at Level 3 as follows. Although the effective channel $\vect{a}_{k}^{\Htran}\vect{g}_{kk}$ is unknown at the CPU, we notice that its average $\vect{a}_{k}^{\Htran} \mathbb{E}\{ \vect{g}_{kk}\}$ is non-zero (e.g., if L-MMSE or MR is used) and deterministic. Therefore, it can be assumed known\footnote{When dealing with ergodic capacities, all deterministic parameters can be assumed known without loss of generality, because these can be estimated using a finite number of transmission resources, while the capacity is only achieved as the amount of transmission resources goes to infinity. Hence, the estimation overhead for obtaining deterministic parameters is  negligible.}
 and used to compute the following achievable SE. 

\begin{proposition} \label{theorem:uplink-capacity-level3}
At Level 3, an achievable SE of UE $k$ is
\begin{equation} \label{eq:uplink-rate-expression-level3}
\begin{split}
\mathacr{SE}_{k}^{(3)} = \left( 1 - \frac{\tau_p}{\tau_c} \right) \log_2  \left( 1 + \mathacr{SINR}_{k}^{(3)}  \right)
\end{split}
\end{equation}
with the effective SINR given by
\begin{equation} \label{eq:uplink-instant-SINR-level3}
\mathacr{SINR}_{k}^{(3)}\! =\!  \frac{ p_{k} \left| \vect{a}_{k}^{\Htran} \mathbb{E}\{ \vect{g}_{kk}\} \right|^2  }{ \!
 \!\sum\limits_{i=1}^{K} p_{i} \mathbb{E} \{  |\vect{a}_{k}^{\Htran}\vect{g}_{ki}|^2  \} - p_{k} \left| \vect{a}_{k}^{\Htran} \mathbb{E}\{ \vect{g}_{kk}\} \right|^2 \!+\! \sigma^2 \vect{a}_{k}^{\Htran}\vect{D}_{k}
 \vect{a}_{k} 
}
\end{equation}
where $\vect{D}_{k}=\diag(\mathbb{E}\{ \|  \vect{v}_{k1}  \|^2\}, \ldots, \mathbb{E}\{ \|  \vect{v}_{kL}  \|^2\}) \in \mathbb{C}^{L \times L}$ and the expectations are with respect to all sources of randomness.
\end{proposition}
\begin{IEEEproof}
The proof is given in Appendix A.
\end{IEEEproof}

The achievable SE above holds for any combining scheme. Particularly, it is valid for both the L-MMSE combining in \eqref{eq:MMSE-combining-single-AP} and the MR combining $\vect{v}_{kl} = \hat{\vect{h}}_{kl}$ that was used in \cite{Nayebi2016a}. Unlike the achievable SE in Proposition~\ref{theorem:uplink-capacity-general}, it holds for any channel estimator (not only for the MMSE estimator \eqref{eq:uplink-rate-expression-general}) but requires channel hardening in order to approximate $\vect{a}_{k}^{\Htran} \vect{g}_{kk}$ with its mean value $\vect{a}_{k}^{\Htran} \mathbb{E}\{ \vect{g}_{kk}\}$. However, this may not occur when the number $N$ of antennas at the APs is relatively small \cite{Chen2018b}. In that case, the SE expression in \eqref{eq:uplink-instant-SINR-level3} underestimates the achievable performance, but is anyway the best available capacity bound.

The structure of \eqref{eq:uplink-instant-SINR-level3} allows computing the deterministic weighting vector  $\vect{a}_{k}$ that maximizes $\mathacr{SINR}_{k}^{(3)}$. This is given as follows.

\begin{corollary} \label{cor:LSFD-optimal}
The effective SINR in \eqref{eq:uplink-instant-SINR-level3} for UE~$k$ is maximized by 
\begin{equation} \label{eq:LSFD-vector}
\vect{a}_{k} =  \left( \sum\limits_{i=1}^{K}
p_{i} \mathbb{E} \{ \vect{g}_{ki} \vect{g}_{ki}^{\Htran} \} + \sigma^2 \vect{D}_{k}
  \right)^{\!\!-1}  \mathbb{E}\{ \vect{g}_{kk}\}
\end{equation}
which leads to the maximum value
\begin{align} \notag
&\mathacr{SINR}_{k}^{(3)} = p_{k}  \mathbb{E}\{ \vect{g}_{kk}^{\Htran}\} \\
& \times \!\! \left( \sum\limits_{i=1}^{K}
p_{i} \mathbb{E} \{ \vect{g}_{ki} \vect{g}_{ki}^{\Htran} \} \!+\! \sigma^2 \vect{D}_{k}
\!-\! p_{k} \mathbb{E}\{ \vect{g}_{kk}\} \mathbb{E}\{ \vect{g}_{kk}^{\Htran}\}
  \right)^{\!\!-1} \!\!\!\mathbb{E}\{ \vect{g}_{kk}\}.
\end{align}
\end{corollary}
\begin{IEEEproof}
It follows from \cite[Lemma B.10]{massivemimobook} by noting that \eqref{eq:uplink-instant-SINR-level3} is a generalized Rayleigh quotient with respect to $\vect{a}_{k}$.
\end{IEEEproof}

Notice that Level 3 is an extension of the LSFD framework in \cite{Adhikary2017a,Nayebi2016a,Trinh2019a,Ngo2018b}, which has previously  been only used in Cell-free mMIMO along with MR combining. In fact, the SE expressions provided in these papers only apply for particular choices of receive combining and not for arbitrary combining as \eqref{eq:uplink-rate-expression-level3}. This makes Proposition~\ref{theorem:uplink-capacity-level3} a novel contribution of this paper.

The signaling required at Level 3 can be quantified as follows. Each AP needs to send $(\tau_c - \tau_p)K$ complex scalars (i.e., $ \check{s}_{kl}$ for all $k$) to the CPU per coherence block. In addition, the computation of \eqref{eq:LSFD-vector} requires knowledge of the $L$-dimensional complex vector $\mathbb{E}\{ \vect{g}_{kk}\}$, of the $L \times L$ Hermitian complex matrix $ \mathbb{E} \{ \vect{g}_{ki} \vect{g}_{ki}^{\Htran} \}$,  and of the real-valued $L \times L$ diagonal matrix $ \vect{D}_{k}$ for $k,i \in \{ 1, \ldots, K\}$. Hence, $KL+(L^2 K^2+KL)/2$ complex scalars are needed in total. These values are summarized in Table~\ref{tab:signaling}. 

\subsection{Level 2: Local Processing \& Simple Centralized Decoding}

Although the optimized LSFD step in Level 3 gives the highest SE among schemes with local combining at each AP, it requires knowledge of a number of statistical parameters that grows quadratically with $L$ and $K$, which can be very large in Cell-free mMIMO. In practice, this large number of parameters need to be jointly estimated by the APs and sent to the CPU. This might not be feasible, especially if the statistics vary with time. To overcome this issue, the CPU can alternatively create its estimate of the signal $s_k$ from UE $k$ by simply taking the average of the local estimates, as proposed in the early papers on the topic \cite{Ngo2017b,Nayebi2017a}.\footnote{Level 2 also includes other cases where the weight $a_{kl}$ is selected based only on the statistical information available at AP $l$. For example, we have tried $a_{kl} = \beta_{lk}^{\nu}$ for different exponents $\nu$ but the performance gap to Level 3 remained to be large. Further research in this direction is needed.} This yields
\begin{align} \label{eq:average-local-estimates}
\hat{s}_k &= \frac{1}{L} \sum_{l=1}^{L} \check{s}_{kl}
\end{align}
where $\check{s}_{kl}$ is given in \eqref{eq:local-data-estimate} and can be obtained by  any local combining vector. Since this is equivalent to setting $\vect{a}_{k} = [ 1/L \, \ldots \, 1/L]^{\Ttran}$ in Proposition~\ref{theorem:uplink-capacity-level3}, the following result is obtained.

\begin{corollary} \label{theorem:uplink-capacity-level2}
At Level 2, an achievable SE of UE $k$ is
\begin{equation} \label{eq:uplink-rate-expression-level2}
\begin{split}
\mathacr{SE}_{k}^{(2)} = \left( 1 - \frac{\tau_p}{\tau_c} \right) \log_2  \left( 1 + \mathacr{SINR}_{k}^{(2)}  \right)
\end{split}
\end{equation}
with the effective SINR given by
\begin{align} \notag
&\mathacr{SINR}_{k}^{(2)} =\\ &\frac{ p_{k} \left|  \sum\limits_{l=1}^{L} \mathbb{E} \left\{ \vect{v}_{kl}^{\Htran} \vect{h}_{kl} \right\} \right|^2  }{ 
\sum\limits_{i=1}^{K} p_{i} \mathbb{E}\!\Big\{ \! \Big|  \sum\limits_{l=1}^{L} \!\vect{v}_{kl}^{\Htran} \vect{h}_{il} \Big|^2 \! \Big\} \!-\! p_{k} \Big|  \sum\limits_{l=1}^{L} \!\mathbb{E} \!\left\{ \vect{v}_{kl}^{\Htran} \vect{h}_{kl} \right\} \! \Big|^2 \!+\!\sigma^2 \sum\limits_{l=1}^{L} \! \mathbb{E}\{ \|  \vect{v}_{kl}  \|^2\}} \label{eq:uplink-instant-SINR-level2}
\end{align}
where the expectations are taken with respect to all sources of randomness.
\end{corollary}

As for Proposition~\ref{theorem:uplink-capacity-level3}, the above SE can be utilized along with any local combining vector and also channel estimator. If MR is used with single-antenna APs (i.e., $N=1$), then Corollary~\ref{theorem:uplink-capacity-level2} reduces to the case considered in \cite{Ngo2017b} and can be computed in closed form (similar results are found in \cite{Buzzi2017a,Zhang2018a,Ozdogan2018a,Fan2019a,Bashar2019a,Yang2019a}). The number of complex scalars to be exchanged per coherence block is the same as at Level 3. The key difference is that no statistical parameters are needed at the CPU. This is summarized in Table~\ref{tab:signaling}.

\subsection{Level 1: Small-Cell Network}
\label{subsec:level1}

The simplest implementation level is when the signal from UE $k$ is decoded by using only the received signal from one AP. In this case, the decoding can be done locally at the AP by using its own local channel estimates without exchange anything with the CPU.\footnote{In all the four levels, the $K$ data streams need to be transmitted to the core network after decoding. This requires a backhaul load proportional to the sum SE, which is not included in Table~\ref{tab:signaling} but is different for each level.} This makes the network truly distributed \cite[Sec.~4.2]{Bjornson2013d} and essentially turns Cell-free mMIMO into a small-cell network. The macro diversity achieved by selecting the best out of many APs could potentially make it competitive compared to conventional Cellular mMIMO with larger cells.

Cell-free mMIMO and small cells were compared in \cite{Ngo2017b,Nayebi2017a} with $N=1$ and an AP selection based on the largest large-scale fading coefficient $\beta_{kl}$. In addition to this, the authors impose that each AP can only serve one UE. Unlike \cite{Ngo2017b,Nayebi2017a}, we remove all these restrictions by assuming an arbitrary number of antennas per AP and letting the AP that gives the highest SE to a specific UE be responsible for decoding its signal. The latter makes the AP association more complex than \cite{Ngo2017b,Nayebi2017a}, but the numerical results in Section~\ref{sec:numerical} show that it vastly improves the performance.\footnote{In practice, selecting the AP that maximizes the SE can be replaced by selecting the AP that maximizes some kind of approximate closed-form SINR. Such a selection rule has the same implementation complexity as selecting the AP with the largest large-scale fading coefficient. However, the challenge is that the SINR is affected by the transmit powers, so if these powers are optimized, the optimization must also involve the AP selection.} Within the above setting, the following result is obtained.

\begin{corollary} \label{theorem:uplink-capacity-general-level1}
At Level 1, an achievable SE of UE $k$ is
\begin{equation} \label{eq:uplink-rate-expression-general-level1}
\begin{split}
\mathacr{SE}_{k}^{(1)} = \left( 1 - \frac{\tau_p}{\tau_c} \right) \max_{l \in \{ 1, \ldots, L \} } \, \mathbb{E} \left\{ \log_2  \left( 1 + \mathacr{SINR}_{kl}^{(1)}  \right) \right\}
\end{split}
\end{equation}
where the instantaneous effective SINR at AP $l$ is
\begin{equation} \label{eq:uplink-instant-SINR_l1}
\!\!\mathacr{SINR}_{kl}^{(1)} \!= \! \frac{ p_{k} |  \vect{v}_{kl}^{\Htran} \hat{\vect{h}}_{kl} |^2  }{ 
 \condSum{i=1}{i \neq k}{K} p_{i} | \vect{v}_{kl}^{\Htran} \hat{\vect{h}}_{il} |^2
+ \vect{v}_{kl}^{\Htran}  \left( \sum\limits_{i=1}^{K} p_{i} \vect{C}_{il} + \sigma^2  \vect{I}_{N} \right)  \vect{v}_{kl}  
}\!\!\!\!\!\!\!\!\!
\end{equation}
and the expectation is with respect to the channel estimates.
The maximum value in \eqref{eq:uplink-instant-SINR_l1} is achieved with the L-MMSE combining in \eqref{eq:MMSE-combining-single-AP} and is given by 
\begin{align}
\!\!\!\!\!\mathacr{SINR}_{kl}^{(1)} =  p_{k}  \hat{\vect{h}}_{kl}^{\Htran} \left( \condSum{i=1}{i \neq k}{K} p_{i}  \hat{\vect{h}}_{il} \hat{\vect{h}}_{il}^{\Htran} + \sum\limits_{i=1}^{K} p_{i} \vect{C}_{il} + \sigma^2  \vect{I}_{N} \!\!\right)^{-1} \!\!\!\!\! \!\!\hat{\vect{h}}_{kl}. \!\!\label{eq:uplink-maximized-SINR-level1}
\end{align}
\end{corollary}
\begin{IEEEproof}
For each AP, the SE is computed in the same way as in Proposition~\ref{theorem:uplink-capacity-general} and the maximum SINR is achieved as in Corollary~\ref{cor:MMSE-combining}.
\end{IEEEproof}

The SE expression above is more general than the one considered for small cells in \cite{Ngo2017b}, where $N=1$ is considered and each AP only estimates the channel of the UE it serves. When considering that special case, the following result is obtained instead.

\begin{proposition} \label{theorem:uplink-capacity-general-level1-MR}
At Level 1 with $N=1$, if AP $l$ decodes the signal from UE $k$ using only its local estimate $ \hat{\vect{h}}_{kl}$, an achievable SE is
\begin{align}  \label{eq:uplink-rate-expression-MR-level1}
\frac{ e^{\frac{1}{\omega_{kl} ( 1 + A_{kl} )} } E_1 \left( \frac{1}{\omega_{kl} ( 1 + A_{kl} )} \right) - e^{\frac{1}{\omega_{kl} A_{kl} }}  E_1 \left( \frac{1}{\omega_{kl} A_{kl}} \right)   }{\ln(2) }
\end{align}
where $A_{kl} = \sum\limits_{i \in \mathcal{P}_k \setminus \{ k \}} \left(\frac{p_{i} \beta_{il}}{p_k \beta_{kl}} \right)^2$ is due to pilot contamination,
\begin{equation}
\omega_{kl} = \frac{p_k^2 \tau_p \beta_{kl}^2}{\Psi_{t_kl} \left(   \sum\limits_{i \not \in \mathcal{P}_k} p_{i} \beta_{il}+  \sum\limits_{i \in \mathcal{P}_k} p_{i} C_{il} + \sigma^2 \right)},
\end{equation}
$E_1(x) = \int_1^{\infty} \frac{e^{-xu}}{ u} du$ denotes the exponential integral, and $\ln(\cdot)$ denotes the natural logarithm.
\end{proposition}
\begin{IEEEproof}
The proof is given in Appendix B.
\end{IEEEproof}

Comparing the achievable SE in \eqref{eq:uplink-rate-expression-MR-level1} with 
\cite[Eq.~(47)]{Ngo2017b} we notice that, despite the different notation, the equivalence only holds when $A_{kl}=0$; that is, in the absence of pilot contamination. 
Although \cite{Ngo2017b} states the result without proof, it seems that the paper has  
neglected the conditioning on the local channel estimate $ \hat{\vect{h}}_{kl}$ when computing the interference power; see \eqref{eq:interference-level1-MR} in Appendix B. This leads to an approximate SE rather than an exact expression. This is why we included Proposition~\ref{theorem:uplink-capacity-general-level1-MR} in this paper and will use it for numerical comparison in Section~\ref{sec:numerical}. 

\begin{remark}
We noticed that the expression in \eqref{eq:uplink-rate-expression-MR-level1} is numerically unstable when $\omega_{kl} ( 1 + A_{kl} )$ and/or $\omega_{kl} A_{kl}$ are small. This is because $e^{1/x} \to \infty $ and $E_1(1/x) \to 0 $ when $x \to 0$. 
When this happens, one can instead utilize the bounds $\frac{x}{1+x} \leq e^{1/x} E_1(1/x) \leq x$ in \cite[Eq.~5.1.19]{Abramowitz} to realize that $ e^{1/x} E_1(1/x) \approx x$ when $x \to 0$.
\end{remark}

\section{Cell-free versus Cellular mMIMO}
\label{sec:numerical}

In this section, we compare the uplink performance of Cell-free mMIMO, with the different cooperation levels and either MR or MMSE/L-MMSE combining, and Cellular mMIMO. We first briefly describe the cellular setup that is considered.

\subsection{Cellular mMIMO Setup}

We consider a cellular network with $L_{\mathrm{c}}=4$ cells, $M_{\mathrm{c}}=100$ antennas per cellular BS, and $K_{\mathrm{c}} = 10$ UEs per cell. 
The block-fading channel from BS $j$ to UE $k$ in cell $l$ is modeled as
\begin{equation} \label{eq:correlated-Rayleigh-model}
\vect{h}_{lk}^{j} \sim \CN \left( \vect{0}, \vect{R}_{lk}^{j}  \right)
\end{equation}
where $\vect{R}_{lk}^{j} \in \mathbb{C}^{M_{\mathrm{c}} \times M_{\mathrm{c}}}$ is the spatial correlation matrix with large-scale fading coefficient $\beta_{lk}^{j} \triangleq \tr(\vect{R}_{lk}^{j})/M_{\mathrm{c}}$ describing the geometric pathloss and shadowing. The uplink transmit power of UE $k$ in cell $l$ is denoted by $p_{lk}\geq 0$.

We assume there are $\tau_p = K_{\mathrm{c}}$ mutually orthogonal pilots and that UE $k$ in every cell uses the same pilot (i.e., pilot reuse one). When using standard MMSE estimation \cite[Th.~3.1]{massivemimobook}, the MMSE estimate of $\vect{h}_{lk}^{j} \in \mathbb{C}^{M_{\mathrm{c}}}$ is given by 
\begin{align} \label{eq:MMSEestimator_h_jli-distribution}
\hat{\vect{h}}_{li}^{j}  \sim \CN \left( \vect{0}, \vect{R}_{li}^{j} -  \vect{C}_{li}^{j}  \right)
\end{align}
and the independent estimation error $\tilde{\vect{h}}_{li}^{j}   \in \mathbb{C}^{M_{\mathrm{c}}}$ is
\begin{align}
\tilde{\vect{h}}_{li}^{j}  \triangleq \vect{h}_{li}^{j}  - \hat{\vect{h}}_{li}^{j}  \sim \CN \left( \vect{0}, \vect{C}_{li}^{j}  \right)
\end{align}
with
\begin{equation}
\!\!\vect{C}_{li}^{j} = \vect{R}_{li}^{j} - p_{li} \tau_p \vect{R}_{li}^{j}
\left( \sum_{l'=1}^{L_{\mathrm{c}}} p_{l'i} \tau_p \vect{R}_{l'i}^{j}  +  \sigma^2  \vect{I}_{M_{\mathrm{c}}} \right)^{-1} \vect{R}_{li}^{j}.\!\!
\end{equation}
An achievable SE of UE $k$ in cell $j$ is \cite[Th.~4.1]{massivemimobook}
\begin{equation}
\begin{split}
\mathacr{SE}^{(\mathrm{c})}_{jk} = \left( 1 - \frac{\tau_p}{\tau_c} \right)  \mathbb{E} \left\{ \log_2  \left( 1 + \mathacr{SINR}_{jk}^{(\mathrm{c})}  \right) \right\}
\end{split}
\end{equation}
where the effective SINR, $\mathacr{SINR}_{jk}^{(\mathrm{c})} $, is maximized by multi-cell MMSE (M-MMSE) combining \cite{BHS18A}. This gives 
\begin{equation}
\begin{split}
&\mathacr{SINR}_{jk}^{(\mathrm{c})} = p_{jk}  (  \hat{\vect{h}}_{jk}^{j} )^{\Htran}  \times\\
&\left( \sum\limits_{l=1}^{L_{\mathrm{c}}} \! \condSum{i=1}{(l,i) \neq (j,k)}{K_{\mathrm{c}}} \! p_{li}  \hat{\vect{h}}_{li}^{j} ( \hat{\vect{h}}_{li}^{j})^{\Htran} + \sum\limits_{l=1}^{L_{\mathrm{c}}} \sum\limits_{i=1}^{K_{\mathrm{c}}} p_{li} \vect{C}_{li}^{j} + \sigma^2  \vect{I}_{M_{\mathrm{c}}} \right)^{-1} \!\!\!\!\!\!\hat{\vect{h}}_{jk}^{j}.
\end{split}
\end{equation}
Other combining schemes can be used but they provide lower SEs. By considering M-MMSE, we thus compare Cell-free mMIMO with the most competitive form of Cellular mMIMO.

\subsection{Simulation Setup and Propagation Model}

The cellular network has 4 square cells in a $1\times1$\,km area, as in Fig.~\ref{fig:illustration_setups}, with 100 co-located antennas per BS. 
The cell-free network is deployed in the same area and has either 400 single-antenna APs (i.e., $N=1$) or 100 four-antenna APs (i.e., $N=4$). Hence, all the network configurations have the same number of antennas. To make a fair comparison, the APs are deployed on a square grid (we consider random deployment later in this section) and the same propagation model is used in all cases. 
We anticipate that the APs in Cell-free mMIMO will be deployed in urban environments with high user loads, roughly 10\,m above the ground. This matches well with the 3GPP Urban Microcell model in \cite[Table~B.1.2.1-1]{LTE2017a} with a 2\,GHz carrier frequency and
\begin{equation}
\beta_{kl} \,  [\textrm{dB}] = -30.5 - 36.7 \log_{10}\left( \frac{d_{kl}}{1\,\textrm{m}} \right)  + F_{kl}
\end{equation}
where $d_{kl}$ is the distance between UE $k$ and AP $l$ (computed as the minimum over different wrap-around cases, and taking the 10\,m height difference into account) and $F_{kl} \sim \mathcal{N}(0,4^2)$ is the shadow fading. The shadowing terms from an AP to different UEs are correlated as \cite[Table B.1.2.2.1-4]{LTE2017a}
\begin{equation} \label{eq:shadowing-decorrelation}
\mathbb{E} \{ F_{kl} F_{ij} \} = 
\begin{cases}
4^2 2^{-\delta_{ki}/9\,\textrm{m}} & l = j \\
0 & l \neq j
\end{cases}
\end{equation}
where $\delta_{ki}$ is the distance between UE $k$ and UE $i$. The second row in \eqref{eq:shadowing-decorrelation} accounts for the correlation of shadowing terms related to two different APs, which is negligible since we have at least 50\,m between adjacent APs in the simulation setup (notice that $2^{-50/9}\approx0.02$).

Since the propagation model from \cite{LTE2017a} is designed for cellular networks, we use the same propagation model for Cellular mMIMO by simply adding an additional index to all the parameters to specify in which cell a particular UE resides. By having a common model for cell-free and cellular networks, we can be sure that the performance differences that we observe are caused by differences in technology characteristics, and not by the propagation model.
There are $K=40$ UEs in the simulation setup, whereof ten are uniformly dropped in each cell and assigned to unique pilots with random indices.\footnote{Each UE in the Cellular mMIMO case is connecting to the BS providing the largest large-scale fading coefficient; that is, $\beta_{jk}^{j} = \max_{l} \beta_{jk}^{j}$.
Due to the shadowing, this might not  be the geographically closest BS.} The same UE locations and pilot assignments are used in the cell-free case, but the shadowing is generated independently.

The cellular BSs and multi-antenna APs are equipped with half-wavelength-spaced uniform linear arrays. The spatial correlation is generated using the Gaussian local scattering model with $15^\circ$ angular standard deviation \cite[Sec.~2.6]{massivemimobook}. All UEs transmit with power $p_k = p_{jk} = 100$\,mW, the bandwidth is 20\,MHz, the noise power is $\sigma^2 = -96$\,dBm, and the coherence blocks contain $\tau_c = 200$ channel uses (e.g., achieved by  2\,ms coherence time and 100\,kHz coherence bandwidth).

\begin{figure} 
        \centering
        \begin{subfigure}[b]{\columnwidth} \centering 
	\begin{overpic}[trim={5mm 0 7mm 0},width=\columnwidth,tics=10]{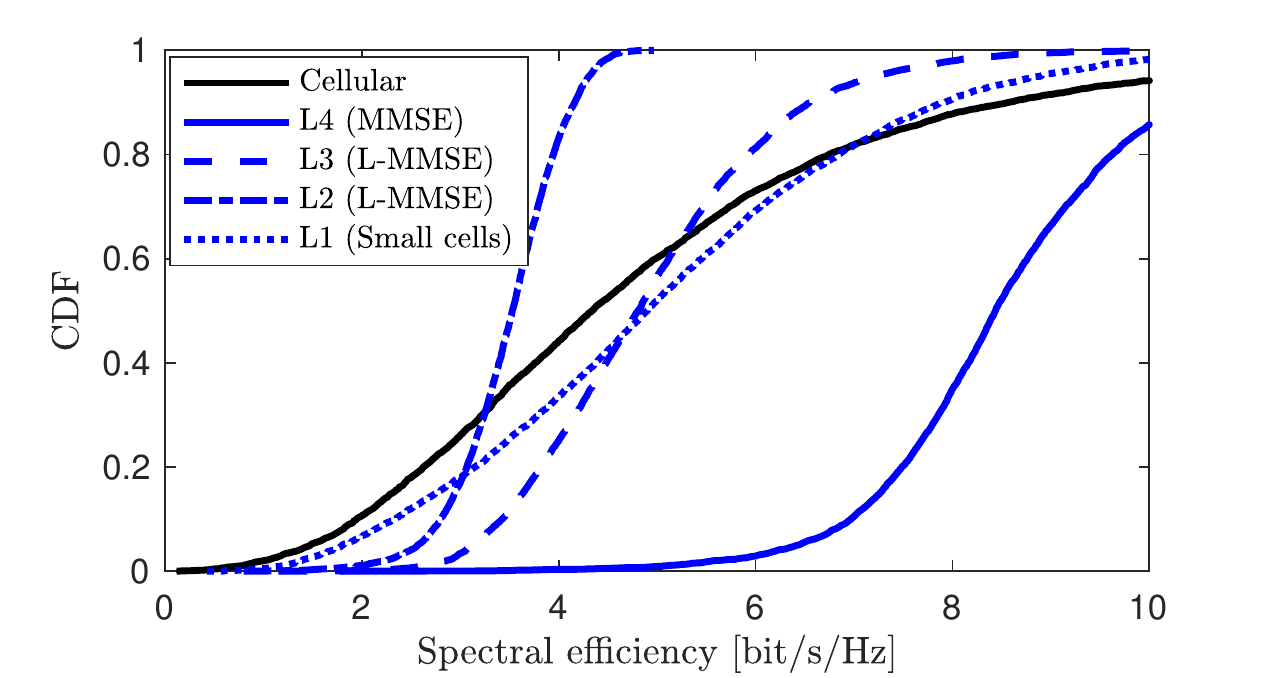}
	\put(12,20){Cellular}
	\put(20,18.5){\vector(1,-1){5}}
	\put(24.5,11.55){$- - - - - - - - - - -\, -$ 90\% likely}
\end{overpic} 
                \caption{Cell-free with $L=400$, $N=1$.} 
                \label{fig:simulationSE_N1}
        \end{subfigure} 
        \begin{subfigure}[b]{\columnwidth} \centering   \vspace{+2mm}
	\begin{overpic}[trim={5mm 0 7mm 0},width=\columnwidth,tics=10]{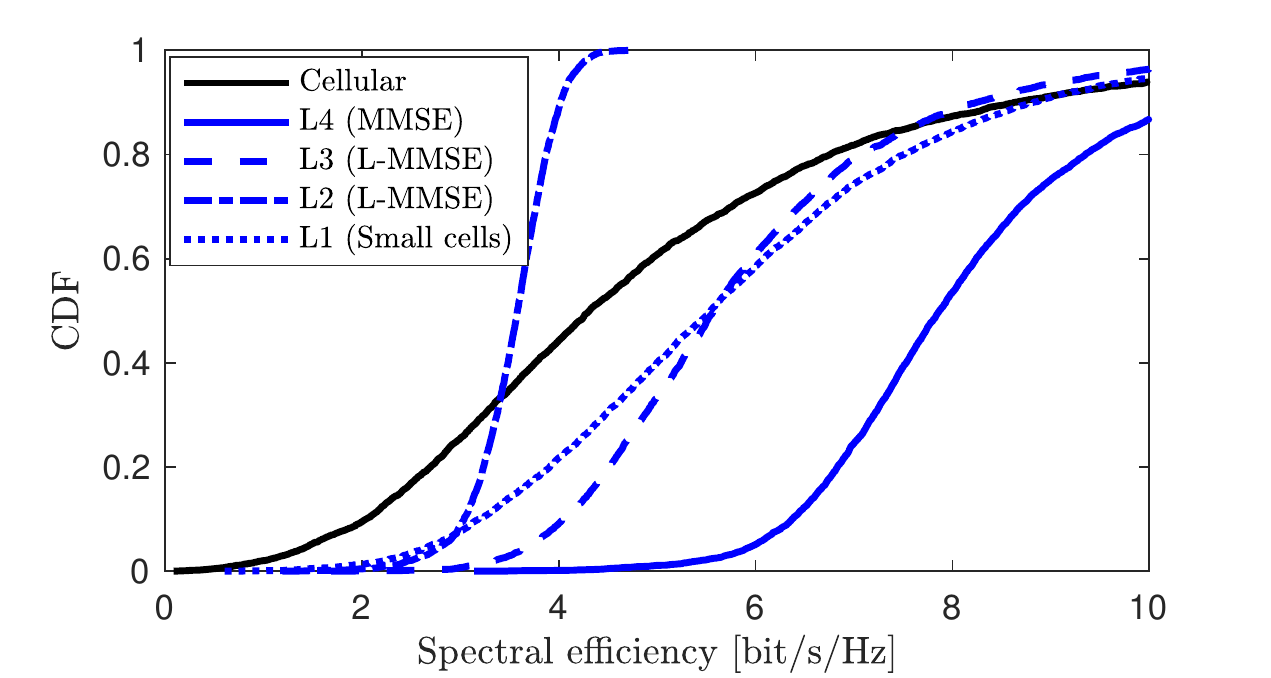}
	\put(12,20){Cellular}
	\put(20,18.5){\vector(1,-1){5}}
		\put(25,11.45){$- - - - - - - - - - -$ 90\% likely}
\end{overpic} 
                \caption{Cell-free with $L=100$, $N=4$.} 
                \label{fig:simulationSE_N4}
        \end{subfigure} 
        \caption{Comparison of Cellular mMIMO and Cell-free mMIMO when using MMSE or L-MMSE combining.}
        \label{fig:simulationSE_N1_N4}  
\end{figure}
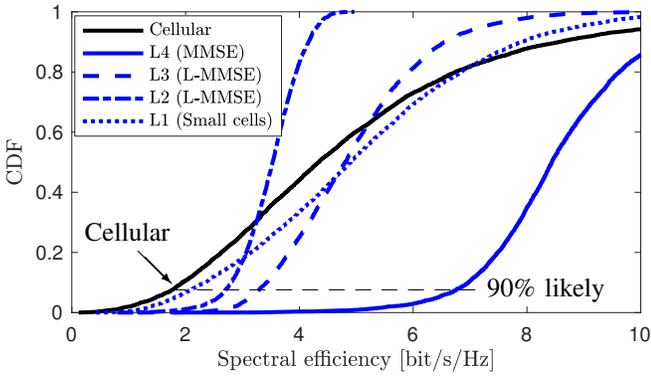
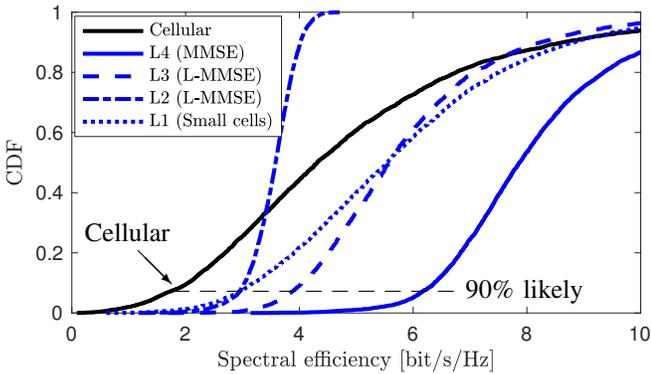

\begin{remark} \label{eq:Ngo-channel-model}
The early Cell-free mMIMO papers \cite{Nayebi2016a,Ngo2017b} used another propagation model, which has since then become standard in the field. However, that model is based on the COST-Hata model from \cite{damosso1999cost} for macro-cells, where the APs are at least 30\,m above the ground and the UEs are at least 1\,km from the AP. This is very different from the micro-cell-like deployment we anticipate for Cell-free mMIMO and it should be noted that the model creators themselves specified that it ``must not be used for micro-cells'' \cite[Ch.~4]{damosso1999cost}. Moreover, the model in \cite{Nayebi2016a,Ngo2017b} has no shadowing when a UE is closer than 50\,m from an AP, which is often the case in Cell-free mMIMO deployments. When the distance is larger, the shadowing decorrelation distance is $10\times$ larger than in the 3GPP model \cite{LTE2017a}. For all these reasons, we believe that the  propagation model used in this paper is a better baseline for evaluating Cell-free mMIMO systems.
\end{remark}

\subsection{Numerical Comparisons} \label{subsec:numerical-comparison}

Fig.~\ref{fig:simulationSE_N1_N4}(a) compares Cellular mMIMO and Cell-free mMIMO with $L=400$ and $N=1$. The figure shows the cumulative distribution function (CDF) of the SE of a randomly located UE, when using MMSE or L-MMSE combining in the cell-free cases. At the 90\% or 95\% likely SE points (i.e., where the vertical axis is 0.1 or 0.05), the cell-free cases perform according to their level: Level 4 provides by far the highest SEs, while Level 1 gives the lowest SEs but is anyway preferable as compared to Cellular mMIMO. Looking at the complete CDF curves, the situation is more complicated since the Level 1 and Cellular mMIMO curves are crossing the Level 2 and Level 3 curves. Hence, UEs with good channel conditions get better performance with these methods. However, Level 4 performs better than Cellular mMIMO for every UE.

Fig.~\ref{fig:simulationSE_N1_N4}(b) considers the same setup but with fewer APs that are equipped with multiple antennas:  $L=100$ and $N=4$. The general trends are the same as in Fig.~\ref{fig:simulationSE_N1_N4}(a)  but Level 4 loses in SE due to the reduced macro diversity; that is, the average distance from a UE to an AP is increased.
In contrast, Level 1 gains in performance since each AP can now suppress interference locally, by using its four antennas. In fact, Level 1 is now comparable to Level 2 for the weakest UEs and substantially better for the strongest UEs.

\begin{figure}[t!]
	\centering \vspace{-1mm}
	\begin{overpic}[trim={5mm 0 7mm 0},width=\columnwidth,tics=10]{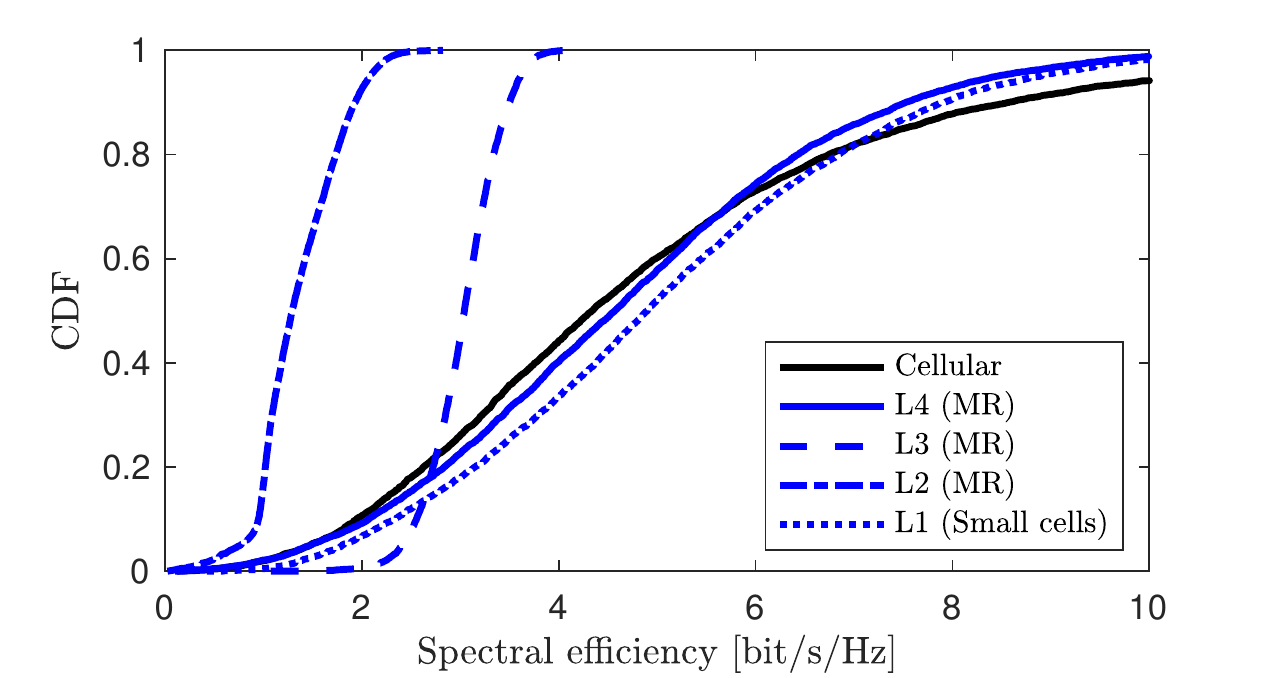}
	\put(75,40){Cellular}
	\put(82,44){\vector(0,1){5}}
	\put(18,11.45){$- - - - -$ 90\% likely}
\end{overpic} \vspace{-3mm}
	\caption{Comparison of Cellular mMIMO with Cell-free ($L=400$, $N=1$) when using MR combining.}
	\label{fig:simulationSE_N1_MR}  \vspace{-3mm}
\end{figure}

Next, Fig.~\ref{fig:simulationSE_N1_MR} considers the case $L=400$, $N=1$ and MR combining, which is the receiver processing advocated in the early papers on Cell-free mMIMO. More precisely, Level 2 was considered in \cite{Ngo2017b} and Level 3  in \cite{Nayebi2016a}. The poor processing leads to a large SE loss, compared to Fig.~\ref{fig:simulationSE_N1_N4}(a), for all levels of Cell-free mMIMO receiver cooperation, except Level 1. In fact, Level 2 is outperformed by both small cells (Level 1) and Cellular mMIMO for every single UE. Note that we are considering single-antenna APs in this figure, so MR processing is suboptimal even in that basic case, and the use of LSFD in Level 3 cannot make up for the performance loss. This is because L-MMSE and MR differ by a non-deterministic scalar and LSFD only involves deterministic scalars. Not even Level 4 performs better than Cellular mMIMO or small cells when using MR, so we can conclude that Cell-free mMIMO should never use the MR scheme.

\subsection{Revisiting ``Cell-free Massive MIMO versus small cells''}

Interestingly, our observations in Fig.~\ref{fig:simulationSE_N1_MR} contradict  the previous results in \cite{Ngo2017b}, where Cell-free mMIMO with `Level 2 (MR)' was shown to perform much better than small cells, in terms of both 95\%-likely and median SE. The reason for the differences is explained in this subsection by reproducing \cite[Fig.~4, Fig.~6]{Ngo2017b} and adding some additional curves to them. The following three-slope propagation model was used in \cite{Ngo2017b}:
\begin{equation}
\beta_{kl} \,  [\textrm{dB}] \!=\! \begin{cases}
-81.2 & d_{kl}< 10\,\textrm{m}\\
-61.2 - 20 \log_{10} \!\left( \frac{d_{kl}}{1\,\textrm{m}} \right)  & 10\,\textrm{m}\leq d_{kl}< 50\,\textrm{m}\\
-35.7 - 35 \log_{10}\!\left( \frac{d_{kl}}{1\,\textrm{m}} \right) \! + F_{kl}\!\!\!\! & d_{kl}\geq 50\,\textrm{m}
\end{cases}
\end{equation}
where $d_{kl}$ is the horizontal distance between UE $k$ and AP $l$ (i.e., ignoring the height difference). The shadowing term $F_{kl}\sim \mathcal{N}(0,8^2)$ only appears when the distance is larger than 50\,m and the terms are correlated as
\begin{equation} \label{eq:shadowing-decorrelation-Ngo}
\mathbb{E} \{ F_{kl} F_{ij} \} = \frac{8^2}{2} \left( 2^{-\delta_{ki}/100\,\textrm{m}} + 2^{-\varrho_{lj}/100\,\textrm{m}} \right)
\end{equation}
where $\delta_{ki}$ is the same as in \eqref{eq:shadowing-decorrelation} and $\varrho_{lj}$ is the distance between AP $l$ and AP $j$.
The maximum UE power is $100$\,mW, the bandwidth is 20\,MHz, the noise power is $\sigma^2 = -92$\,dBm, and the coherence blocks are determined by $\tau_c = 200$.

\begin{figure} 
        \centering
        \begin{subfigure}[b]{\columnwidth} \centering 
	\begin{overpic}[trim={5mm 0 7mm 0},width=\columnwidth,tics=10]{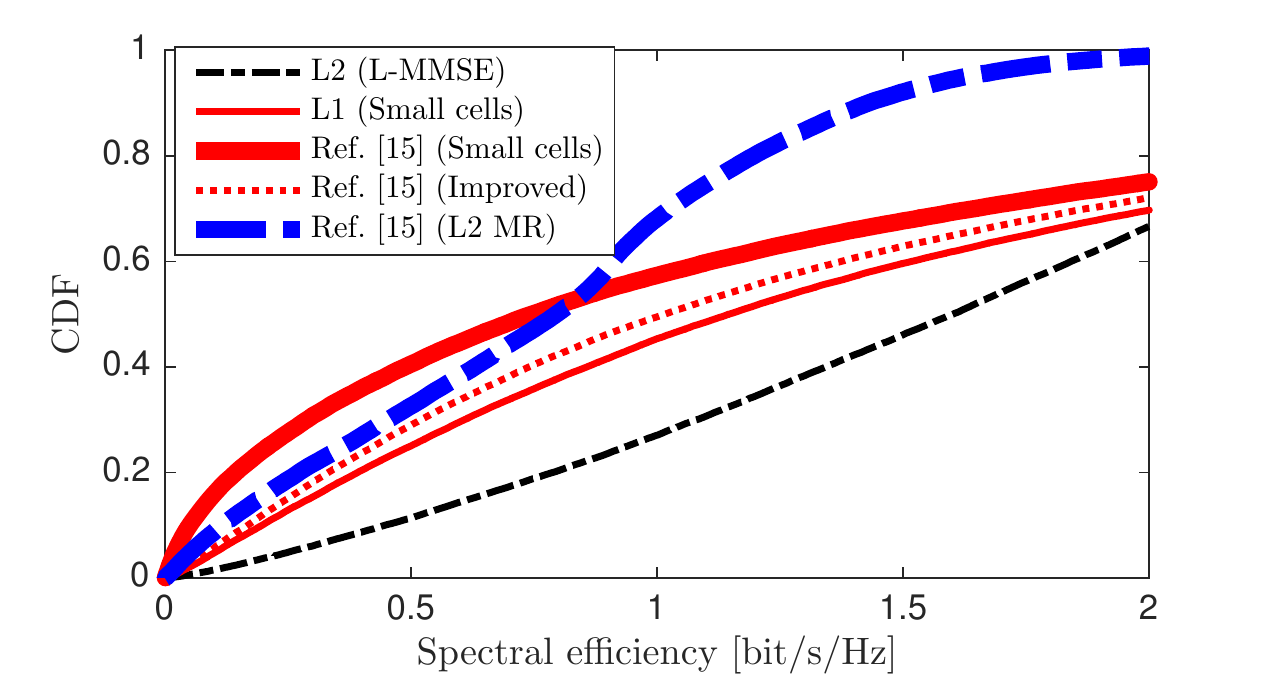}
\end{overpic} 
                \caption{Full power.} 
                \label{simulationFigure4_full}
        \end{subfigure} 
        \begin{subfigure}[b]{\columnwidth} \centering   \vspace{+2mm}
	\begin{overpic}[trim={5mm 0 7mm 0},width=\columnwidth,tics=10]{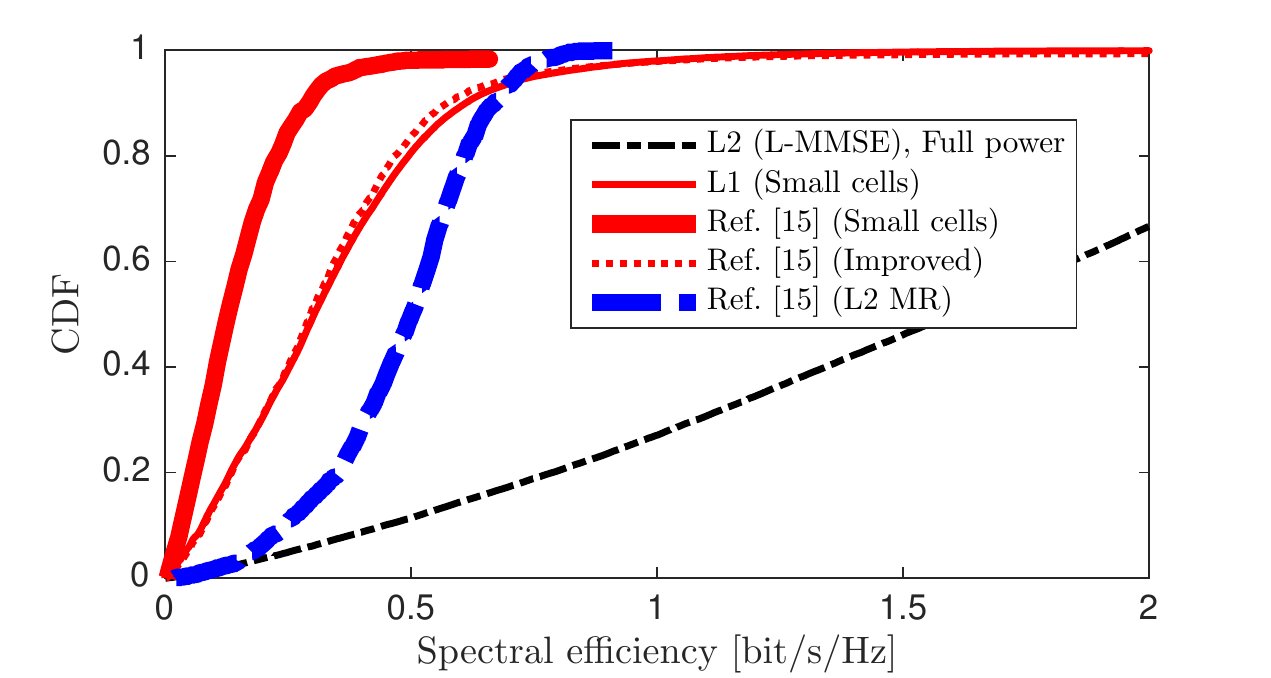}
\end{overpic} 
                \caption{Max-min power control.} 
                \label{simulationFigure4_maxmin}
        \end{subfigure} 
        \caption{Comparison of Cell-free mMIMO at Level 2 and small cells, using different SE expressions and AP assignments. The UEs either transmit with full power or optimizes the power as described in \cite{Ngo2017b}.  The thick lines correspond to the curves in \cite[Fig.~4, Fig.~6]{Ngo2017b}.}
        \label{simulationSE_threeslope}  
\end{figure}

We consider the same setup as in \cite{Ngo2017b} with $L=100$ uniformly distributed APs in a $1\times1$\,km area, $N=1$ antenna per AP, $K=40$ uniformly distributed UEs, and $\tau_p = 20$ orthogonal pilots. The pilots are assigned to the UEs according to the greedy algorithm described in \cite[Sec.~IV.A]{Ngo2017b}, with the only difference that we use the uplink SE as the metric in Step 2 of the algorithm (instead of the downlink SE). Moreover, we use Proposition~\ref{theorem:uplink-capacity-general-level1-MR} to accurately compute the SE with small cells, but this has little impact on the results. The thick lines in Fig.~\ref{simulationSE_threeslope} correspond to the original curves from \cite{Ngo2017b} with correlated shadowing.  Fig.~\ref{simulationSE_threeslope}(a) considers the case when the UEs transmit at full power (as in  \cite[Fig.~6]{Ngo2017b}) and Fig.~\ref{simulationSE_threeslope}(b) considers the case when the UEs transmit pilots at full power but reduce the power during the data transmission to optimize the network-wide max-min fairness (as in  \cite[Fig.~4]{Ngo2017b}). To this end, we use the same optimization algorithms as in \cite{Ngo2017b}.

In the full power case, in Fig.~\ref{simulationSE_threeslope}(a), it is clear from the thick curves that Cell-free mMIMO at Level 2 with MR gives the UEs with the  50\% worst channel conditions substantially higher SE than with small cells. The remaining UEs get better SEs with small cells, which indicates that the considered Cell-free system is not well implemented---since the cell-free network has access to more APs and signal observations when decoding a UE's signal, the performance should be better for \emph{everyone}. Moreover, the comparison in \cite[Fig.~6]{Ngo2017b} is based on a suboptimal assignment of UEs to small cells; the UEs are sequentially selecting the AP that has the largest large-scale fading coefficient $\beta_{kl}$ (i.e., the best channel), but only among those that are not already serving another UE. If we change that to let each UE being served by the AP giving the highest SE, represented by the curve `Ref.~[14] (Improved)', then the performance gap between Cell-free mMIMO and small cells diminishes. The reason is that around 40\% of the UEs prefer to be served by another small cell. Additionally,
if we use the new improved SE expression in Corollary~\ref{theorem:uplink-capacity-general-level1}, represented by the curve `L1 (Small cells)', all the UEs get higher SE with small cells than with Cell-free mMIMO.

Does this mean that small cells are actually better than Cell-free mMIMO? The answer is no. Indeed, as observed in the last subsection, the problem is that MR combining  performs badly in Cell-free mMIMO, even if single-antenna APs are used. By simply changing to Level 2 with L-MMSE combining, the rightmost curve in 
Fig.~\ref{simulationSE_threeslope}(a) is achieved, which gives uniformly higher SE to all the UEs than when using small cells. Even higher SE can be achieved by considering Level 3 or Level 4 implementations.

The results in Fig.~\ref{simulationSE_threeslope}(b) with max-min power control are different and more in line with the observations made in \cite{Ngo2017b}: Level 2 with MR gives much higher SE than small cells, but the gap can be reduced by selecting APs based on the maximum SE rather than the maximum $\beta_{kl}$ (represented by the curves `Ref. [15] (Improved)' and `L1 (Small cells)'). The benefit of max-min power control can be seen by considering the two thick lines (obtained as in \cite{Ngo2017b}):  the lower end of the CDF curves are shifted to the right as compared to the full power case in Fig.~\ref{simulationSE_threeslope}(a), yielding a higher guaranteed SE level. Nevertheless, the use of L-MMSE combining is more appealing than the use of max-min power control, as can be seen from the rightmost curve in Fig.~\ref{simulationSE_threeslope}(b) that considers Level 2 with L-MMSE and full power transmission. This approach gives the same performance as `L2 MR' with max-min fairness for the 2\% weakest UEs, but higher SE for all other UEs; for example, it achieves a 40\% higher 95\%-likely SE and a $3\times$ higher median SE. Hence, if  L-MMSE processing is used, advanced power control is not needed in Cell-free mMIMO to give good performance to the weakest UEs.

\begin{figure}[t!]
	\centering \vspace{-1mm}
	\begin{overpic}[trim={5mm 0 7mm 0},width=\columnwidth,tics=10]{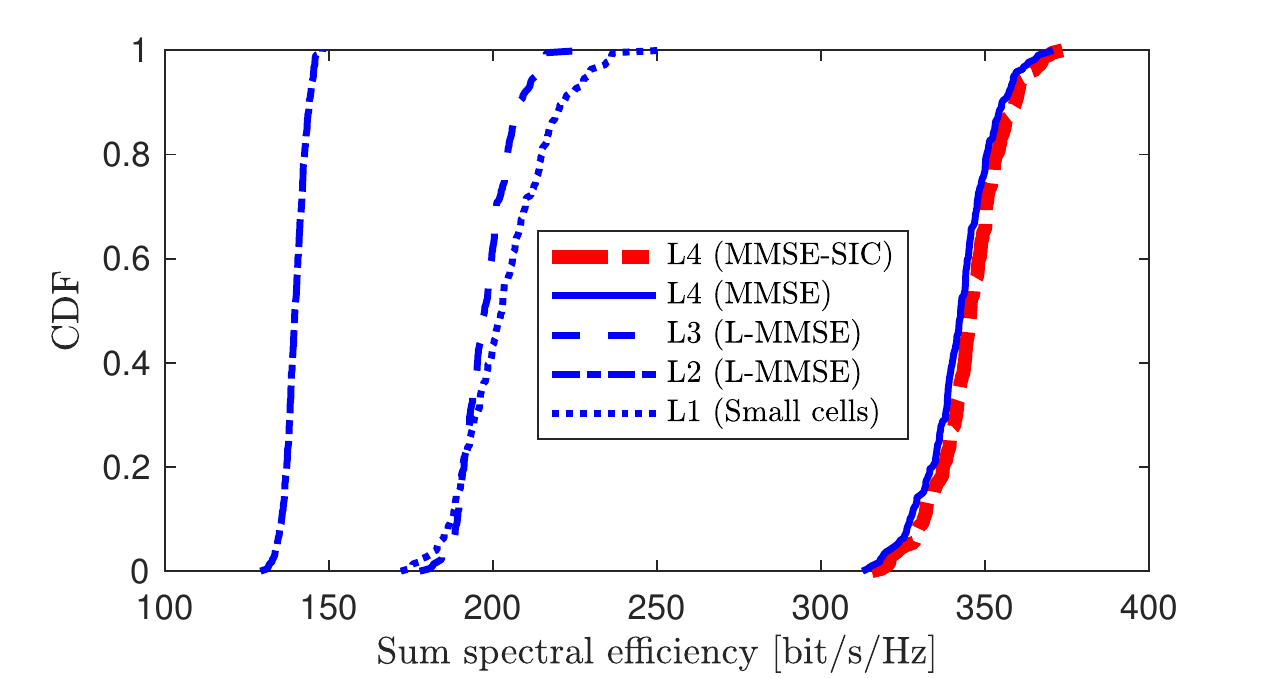}
\end{overpic} \vspace{-3mm}
	\caption{CDF of the sum SE over different random user locations with $L=100$, $N=4$, $K=40$, $\tau_p=10$. The four cooperation levels are compared with MMSE-SIC, based on Proposition \ref{prop:MMSE-SIC}.}
	\label{fig:simulationSE_N1_SIC}  \vspace{-3mm}
\end{figure}

\section{Level 4 With Non-linear Decoding}\label{sec:MMSE-SIC}

Section~\ref{subsec:numerical-comparison} showed that Level 4 can provide vastly higher SE than the other cooperation levels in Cell-free mMIMO. The comparison is based on using linear receive combining, but another benefit of centralizing the signal processing at a CPU is that more advanced decoding methods can potentially be used, since network-wide CSI and high computational resources are available. In this section, we investigate the potential benefits of the non-linear successive interference cancelation (SIC) method \cite[Sec.~8.3.4]{Tse2005a} in Cell-free mMIMO, which means that the CPU decodes one UE signal at a time, and then sequentially subtracts interference that the decoded signal caused to the remaining signals. The interference cannot be fully canceled since the CPU has imperfect CSI, but it can still improve the SE of the UEs compared to linear combining.

\begin{proposition} \label{prop:MMSE-SIC}
At Level 4, if the MMSE estimator is used to compute channel estimates for all UEs and the signals are decoded using MMSE combining and SIC (MMSE-SIC), then for any decoding order an achievable sum SE is
\begin{equation} \label{eq:uplink-rate-expression-general-SIC}
\mathacr{SSE}^{\mathrm{(SIC)}} = \left( 1 - \frac{\tau_p}{\tau_c} \right) \mathbb{E} \left\{ \log_2 \det \left( \vect{I}_K + \vect{P} \hat{\vect{H}}^{\Htran} \vect{E}^{-1} \hat{\vect{H}} \right) \right\}
\end{equation}
where $\vect{P} = \diag(p_1,\ldots,p_K)$, $\hat{\vect{H}} = [ \hat{\vect{h}}_{1} \, \ldots \,  \hat{\vect{h}}_{K}] \in \mathbb{C}^{LN \times K}$, $\vect{E} = \sum_{i=1}^{K} p_{i} \vect{C}_{i} + \sigma^2  \vect{I}_{LN}$, and the expectation is with respect to the channel estimates.
\end{proposition}
\begin{IEEEproof}
The proof is given in Appendix C.
\end{IEEEproof}

Proposition~\ref{prop:MMSE-SIC} provides the sum SE of the Cell-free mMIMO network, and not the individual SEs of the UEs. The reason is that the latter depends on the decoding order; that is, the later a UE is decoded, the less interference it will be affected by and thereby it will gain more in SE compared to using linear combining. Irrespective of the decoding order, all UEs get at least as high SE with MMSE-SIC as with MMSE combining.

Fig.~\ref{fig:simulationSE_N1_SIC} revisits the scenario in Fig.~\ref{fig:simulationSE_N1_N4}(a) by considering Cell-free mMIMO with $L=400$ and $N=1$. The CDF of the sum SE over different random realizations of the UE locations is plotted when using either Levels 1-4 with MMSE/L-MMSE combining or Level 4 with MMSE-SIC, based on Proposition \ref{prop:MMSE-SIC}. The MMSE-SIC method improves the sum SE, but the average gain over `L4 (MMSE)' is only 1\%. The reason for such modest gain is the favorable propagation phenomenon that makes the UEs' channels nearly orthogonal \cite{Chen2018b}, meaning that the inter-user interference is effectively canceled by the MMSE processing described in Section~\ref{sec:four-level-receiver-cooperation}. Hence, we conclude that non-linear processing is not needed in Cell-free mMIMO. This is also the reason why we did not present the detailed per-user SEs in this section.

Another observation that can be made from Fig.~\ref{fig:simulationSE_N1_SIC} is that Level 1 and Level 3 provide roughly the same sum SE, while Level 2 is far behind in performance. The large gap to Level 4 further reinforces the point that a centralized implementation is strongly preferred in Cell-free mMIMO.

\section{A Look at the Fronthaul Signaling Load}
\label{sec:fronthaul}

The reported results show that a Level 4 implementation is strongly preferred.
The counterargument might be that such an implementation would require much more fronthaul signaling than Level 2 and Level 3, but we will now show that it is not necessarily the case. By using the formulas in Table~\ref{tab:signaling}, Level 4 requires less signaling if
\begin{equation} \label{eq:ratio-signaling}
\frac{\tau_c NL}{(\tau_c-\tau_p) KL} = \frac{\tau_c}{\tau_c-\tau_p} \frac{N}{K}  < 1.
\end{equation}
Since $\frac{\tau_c}{\tau_c-\tau_p}  \approx 1$ and $K \gg N$ are typical for Cell-free mMIMO, Level 4 actually requires \emph{much less} signaling.

\begin{figure}[t!]
	\centering \vspace{-1mm}
	\begin{overpic}[trim={5mm 0 7mm 0},width=\columnwidth,tics=10]{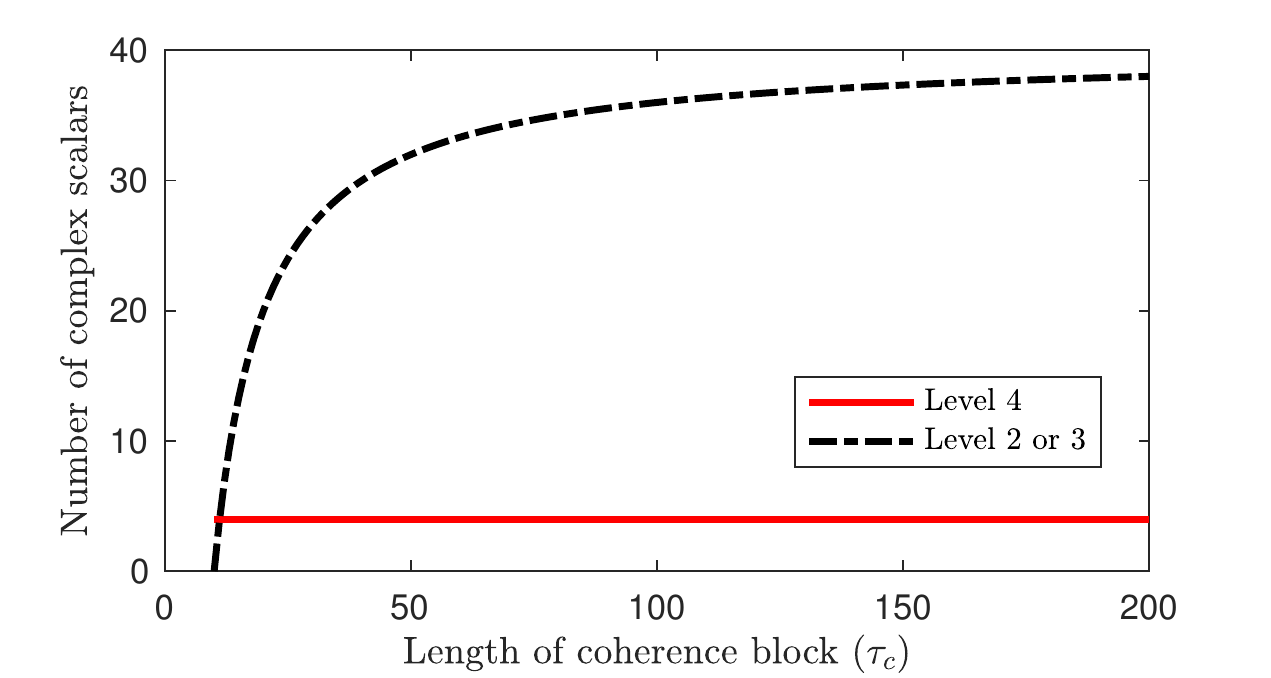}
\end{overpic} \vspace{-3mm}
	\caption{Number of complex scalars that needs to be shared between an AP and the CPU per channel use ($L=100$, $N=4$, $K=40$, $\tau_p=10$).}
	\label{fig:simulationSignaling}  \vspace{-3mm}
\end{figure}

Fig.~\ref{fig:simulationSignaling} shows how many complex scalars need to be sent from an AP to the CPU per channel use, as a function of the coherence block length $\tau_c$. We consider the same setup as in Fig.~\ref{fig:simulationSE_N1_N4}(b):  $L=100$, $N=4$, $K=40$, and $\tau_p=10$.
Level 4 requires more signaling if $\tau_c \leq 11$, while much less signaling is required when $\tau_c$ becomes a hundred, as in practical systems. As $\tau_c \to \infty$, Level 2 and 3 require $K/N=10$ times more fronthaul signaling than Level 4.
The reason is that the received data signals constitute a much larger number of scalars than the  channel estimates. Since $K \geq N$ is typically the case in Cell-free mMIMO, Level 2 and Level 3 increase the fronthaul signaling by processing the $N$-dimensional vector $\vect{y}_{l}$ into the $K$-dimensional vector $[\check{s}_{1l},\ldots\check{s}_{Kl}]^{\Ttran}$. In practice, an AP will not serve all the UEs in the network but only those with a good channel. Nevertheless, as long as each AP serves more UEs than it has antennas (e.g., more than one UE in conventional Cell-free mMIMO with $N=1$), Level 4 is preferable in terms of fronthaul signaling.

Admittedly, this comparison assumes that all scalars are shared with infinite precision, while in practice it is plausible that the pilot signals require higher bit-resolution when sent to the CPU than the data signals. On the other hand, the pilot signals constitute only a minor fraction of the total signaling and \cite{Bashar2018a,Maryopi2019a} recently showed that the estimates can be compressed rather well.

\subsection{Serial Fronthaul}

Since Level 2 and Level 3 necessarily provide lower SE than Level 4, these levels are only  practically interesting if they require lower fronthaul capacity. 
The previous example shows that this is not the case when each AP transmits individually over the fronthaul, but there are alternative solutions that reduce the fronthaul capacity requirement. In particular, this happens when several APs are deployed along the same wired connection, as illustrated in the lower right corner of Fig.~\ref{fig:illustration_setups}(b). 

Suppose AP 1 and AP 2 share a fronthaul connection in this way.  When the locally estimated signal $\check{s}_{k1}$ at AP $1$ is sent over the fronthaul to AP $2$, this AP will compute $\check{s}_{k1} + \check{s}_{k2}$. The result is then sent to the CPU, which can still form its signal estimate in \eqref{eq:average-local-estimates} at Level 2, since it is the summation of the local estimates at all the APs. By instead transmitting the weighted local estimates $a_{kl}\check{s}_{kl} $ over the fronthaul, Level 3 can be implemented in the same sequential fashion (assuming that $a_{kl}$ can be computed locally at AP $l$).

Since only one scalar per UE is transmitted over each segment of the fronthaul, the capacity requirement does not grow with the number of APs that are sharing the wired connection. In the extreme case when all APs are deployed along the same wire, the number of complex scalars sent over the fronthaul per coherence block reduces from $(\tau_c-\tau_p) KL $ in Table~\ref{tab:signaling} to $(\tau_c-\tau_p) K$. This type of serial fronthaul is needed for Level 2 and Level 3 to make practical sense, which is why it is adopted by the radio stripes concept described in \cite{Interdonato2018}.

\section{Conclusions}
\label{sec:conclusion}

This paper introduced a taxonomy for Cell-free mMIMO with four different implementation levels, from fully centralized to fully distributed, and generalized previous results to account for multi-antenna APs, spatially correlated fading, and arbitrary receive combining. The majority of previous papers on this topic relied upon a distributed implementation with local MR processing. Remarkably, we discovered that this is basically the worst way to operate cell-free networks.

Firstly, local MMSE processing provides substantially higher SE than MR, and is the key prerequisite for Cell-free mMIMO to outperform conventional Cellular mMIMO and small-cell networks. Importantly, this is the case even if each AP is equipped with only one antenna; local MMSE processing can roughly double the SE per UE.

Secondly, we showed that a centralized implementation, with all the signal processing taking place at an edge-cloud processor (a.k.a.~CPU in the cell-free literature), is highly preferable compared to distributed alternatives. In fact, the centralized Level 4 implementation can simultaneously increase the SE and reduce the fronthaul signaling. Linear processing is sufficient at Level 4 since non-linear processing provides negligible gains due to the favorable propagation property \cite{Chen2018b}.
The pCell technology \cite{Perlman2015a} is an example of a centralized cell-free system, which demonstrates that it is practically feasible.
A serial fronthaul is needed to make a distributed implementation competitive in terms of the fronthaul capacity requirements, and an improved version of Level 3 needs to be developed to reduce the performance gap to Level 4. Non-linear processing can be useful at Level 3 and the compute-and-forward relaying framework can potential guide the development of such methods \cite{Nazer2011a,Park2013a,Zhou2016a}.

An interesting analogy can be made between the results in this paper and recent developments in the Cellular mMIMO area. The seminal paper \cite{Marzetta2010a} advocated for using MR processing, based on asymptotic arguments. MR was known to be suboptimal when having a small number of antennas, but anyway became the most well-studied method in the literature since the SE can be computed in closed-form, even with more complicated system models containing spatially correlated fading and/or hardware impairments \cite{massivemimobook}. However, recent works have shown that M-MMSE processing greatly outperforms MR even in the asymptotic regime \cite{BHS18A}.
Similarly, the main conclusion of this paper is that it is time to forget about MR also in Cell-free mMIMO and instead consider only MMSE-based schemes---irrespective of the level of cooperation among the APs and the number of antennas used at each one.

\appendices

\section*{Appendix A \\ Proof of Proposition~\ref{theorem:uplink-capacity-level3}}

Since the CPU does not have knowledge of the channel estimates, it needs to treat the average channel gain $\vect{a}_{k}^{\Htran}\mathbb{E}\{\vect{g}_{kk}\}$ as the true deterministic channel. Hence, the signal model is
\begin{equation} \label{eq:CPU-decoding}
\hat{s}_k = \vect{a}_{k}^{\Htran}\mathbb{E}\{\vect{g}_{kk}\} s_k + \upsilon_k
\end{equation}
which is a ``deterministic'' channel with the additive interference plus noise term
\begin{align} \notag \upsilon_k 
=  \big(\vect{a}_{k}^{\Htran}\vect{g}_{kk} &-  \vect{a}_{k}^{\Htran} \mathbb{E}\{ \vect{g}_{kk}\}\big)s_k\\&+ \!
\sum\limits_{i=1,i\ne k}^{K} \vect{a}_{k}^{\Htran}  \vect{g}_{ki} \vect{g}_{ki}^{\Htran} \vect{a}_{k} s_i + \! \vect{n}^\prime_{k}.
\end{align}
The interference term $\upsilon_k$ has zero mean and is uncorrelated with the signal term in \eqref{eq:CPU-decoding} since
\begin{align}
\underbrace{\mathbb{E}\{ \vect{a}_{k}^{\Htran}\vect{g}_{kk} -  \vect{a}_{k}^{\Htran} \mathbb{E}\{ \vect{g}_{kk}\}\}}_{=0}\mathbb{E}\{|s_k|^2\} = 0.
\end{align}
Therefore, we can apply \cite[Cor.~1.3]{massivemimobook} to obtain the achievable SE in \eqref{eq:uplink-rate-expression-level3}.

\section*{Appendix B\\Proof of Proposition~\ref{theorem:uplink-capacity-general-level1-MR}}

In this proof, we drop the bold face to emphasize that all parameters are scalars.
Using the capacity lower bound in \cite[Cor.~1.3]{massivemimobook} with $\hat{h}_{kl}$ as the known channel realization, an achievable SE is 
\begin{align} \label{eq:SE_level1_MR}
 \mathbb{E} \left\{ \log_2 \left( 1 + \frac{ p_{k} |  \hat{h}_{kl} |^2  }{ \mathbb{E} \{ | \upsilon |^2 \, |  \hat{h}_{kl} \} + \sigma^2 } \right) \right\}
\end{align}
where $\upsilon =  \tilde{h}_{kl} s_k + \sum_{i \neq k} h_{il} s_i$ and 
\begin{align} \label{eq:interference-level1-MR}
\mathbb{E} \{ | \upsilon |^2 \, |  \hat{h}_{kl} \} = \!\!\sum\limits_{i \in \mathcal{P}_k \setminus \{ k\}} \frac{p_{i}^2 \beta_{il}^2}{p_k \beta_{kl}^2} |  \hat{h}_{kl} |^2 +
\sum\limits_{i \not \in \mathcal{P}_k} p_{i} \beta_{il}+  \sum\limits_{i \in \mathcal{P}_k} p_{i} C_{il}
\end{align}
by exploiting the fact that $\hat{h}_{il}$ and $\hat{h}_{kl}$ are independent for all $i \not \in \mathcal{P}_k$ and $\hat{h}_{il} = \frac{\sqrt{p_i} \beta_{il} }{\sqrt{p_k} \beta_{kl} } \hat{h}_{kl}$ for all $i \in \mathcal{P}_k$.
By inserting \eqref{eq:interference-level1-MR} into \eqref{eq:SE_level1_MR}, we can expand the expression as
\begin{align} \notag
& \mathbb{E} \left\{ \log_2 \left( 1 +  |  \hat{h}_{kl} |^2 \frac{ p_k \left( 1 + A_{lk} \right) }{ \sum\limits_{i \not \in \mathcal{P}_k} p_{i} \beta_{il}+  \sum\limits_{i \in \mathcal{P}_k} p_{i} C_{il} + \sigma^2} \right) \right\}
\\ &\quad  -  \mathbb{E} \left\{ \log_2 \left( 1 +  |  \hat{h}_{kl} |^2 \frac{ p_k A_{lk} }{ \sum\limits_{i \not \in \mathcal{P}_k} p_{i} \beta_{il}+  \sum\limits_{i \in \mathcal{P}_k} p_{i} C_{il} + \sigma^2} \right) \right\} 
\end{align}
and compute each of the expectations using \cite[Lemma~3]{Bjornson2010c} and 
$ \hat{h}_{kl} \sim \CN (0, p_k \tau_p \beta_{kl}^2 /\Psi_{t_k l} )$ to obtain the final expression in \eqref{eq:uplink-rate-expression-MR-level1}.

\section*{Appendix C\\Proof of Proposition~\ref{prop:MMSE-SIC}}

The received signal in \eqref{eq:received-data-central2} for Level 4 can be expressed as
\begin{align} 
\vect{y} =\sum_{i=1}^{K} \hat{\vect{h}}_{i} s_i + \vect{e} \label{eq:received-data-central-SIC}
\end{align}
where $\vect{e} \triangleq \vect{n} +  \sum_{i=1}^{K} \tilde{\vect{h}}_{i} s_i$ has zero mean and correlation matrix $\vect{E}$. Since the MMSE channel estimates are known and  $\vect{e}$ is uncorrelated with $\hat{\vect{h}}_{i} s_i$ for all $i$, \eqref{eq:received-data-central-SIC} can be treated as a multiple access channel with colored noise.
In the worst case, in terms of mutual information, the colored noise is independent of the desired signals and Gaussian distributed. Hence, we can apply pre-whitening followed by standard results on MMSE-SIC receivers to obtain the achievable sum SE \cite[Sec.~8.3.4]{Tse2005a}
$\mathbb{E} \{ \log_2 \det ( \vect{I}_{NM} +  \vect{A}^{-1/2}  \hat{\vect{H}} \vect{P}\hat{\vect{H}}^{\Htran}  \vect{A}^{-1/2}   ) \}$. This expression reduces to \eqref{eq:uplink-rate-expression-general-SIC} by utilizing the fact that $\det(\vect{I} + \vect{B} \vect{C}) = \det(\vect{I} + \vect{C}\vect{B} )$ for any matrices $\vect{B},\vect{C}$ of compatible sizes, and by including the pre-log factor $1 - \tau_p /\tau_c$ that is the fraction of channel uses used for data. Note that the sum SE expression is independent of the decoding order.

\bibliographystyle{IEEEtran}
\bibliography{IEEEabrv,refs}

\end{document}